\documentclass[Times,twocolumn,tighten]{aastex63}
\usepackage{graphicx}
\usepackage{multirow}
\usepackage{txfonts}
\usepackage{natbib}
\usepackage{url}
\usepackage{hyperref}

\shorttitle{X-ray Properties of MAXI J0637-430}
\shortauthors{K. Chatterjee et al.}

\begin{document}

\title{MAXI J0637-430: A Possible Candidate for Bulk Motion Comptonization?}

\correspondingauthor{Kaushik Chatterjee}
\email{mails.kc.physics@gmail.com}

\author[0000-0002-6252-3750]{Kaushik Chatterjee}
\affiliation{South Western Institute for Astronomical Research, Yunnan University, University Town, Chenggong, Kunming 650500, P. R. China}
\affiliation{Institute of Astronomy Space and Earth Science, AJ 316, Sector II, Salt Lake, Kolkata 700091, India}
\affiliation{Institute of Astronomy, National Tsing Hua University, Hsinchu 300044, Taiwan}

\author[0000-0003-1856-5504]{Dipak Debnath}
\affiliation{Institute of Astronomy Space and Earth Science, AJ 316, Sector II, Salt Lake, Kolkata 700091, India}
\affiliation{Institute of Astronomy, National Tsing Hua University, Hsinchu 300044, Taiwan}

\author[0000-0002-6640-0301]{Sujoy Kumar Nath}
\affiliation{Indian Centre for Space Physics, 43 Chalantika, Garia St. Road, Kolkata 700084, India}

\author[0000-0002-5617-3117]{Hsiang-Kuang Chang}
\affiliation{Institute of Astronomy, National Tsing Hua University, Hsinchu 300044, Taiwan}


\begin{abstract}

The transient Galactic black hole candidate MAXI J0637-430 went through an outburst in 2019--20 for the very first time. This outburst was active for almost 
6 months from November 2019 to May 2020. We study the spectral properties of this source during that outburst using archival data from NICER, Swift, and NuSTAR 
satellites/instruments. We have analyzed the source during 6 epochs on which simultaneous NICER--NuSTAR and Swift/XRT--NuSTAR data were available. Using both 
phenomenological and physical model fitting approaches, we analyzed the spectral data in the broad $0.7-70$ keV energy band. We first used a combination of disk 
blackbody with power-law, disk blackbody with broken power-law, and disk blackbody with power-law and bmc models. For a better understanding of the accretion 
picture, e.g., understanding how the accretion rates change with the changing size of the perceived Compton cloud, we used the two-component advective 
flow (TCAF) model with broken power-law, TCAF with power-law and bmc models. For last 3 epochs, the diskbb+power-law and TCAF models were able to spectrally 
fit the data for acceptable $\chi^2/DOF$. However, for the first 3 epochs, we needed an additional component to fit spectra for acceptable $\chi^2/DOF$. From our
analysis, we reported about the possible presence of another component during these first 3 epochs when the source was in the high soft state. This additional
component in this state is best described by the bulk motion Comptonization phenomenon. From the TCAF model fitting, we estimated the average mass of the source as 
$8.1^{+1.3}_{-2.7}~M_\odot$.

\end{abstract}


\section{Introduction}

X-ray binaries (XRBs) are quite common in astronomical binary systems. These systems are important to study as they are characterized by accretion as the power 
source  (Frank, King \& Raine 2002). XRBs mainly consist of a compact object as the primary star and a companion as the donor star. The compact object could be 
either a black hole or a neutron star, both of which are the end products of stellar bodies. There are several ways to classify XRBs. Depending on the 
mass of the companion, they are mainly of two types (Remillard \& McClintock 2006): high-mass X-ray binaries (HMXRBs) and low-mass X-ray binaries (LMXRBs). 
Depending on the nature of the variability of outbursts, XRBs are also classified into two types: transient and persistent sources. Flux or counts of 
the persistent sources stay higher than just the detection level most of the time ($L > 10^{36}$ erg/s; Chen et al. 1997), while the flux of transient 
sources occasionally goes above the detection level and mostly stays in the dormant state ($L < 10^{32}$ ~erg/s; Hannikainen et al. 2005), known as 
quiescence. These transients go for outbursts, which last from weeks to months (Tetarenko et al. 2016). As of today, most of the observed transients 
have been classified as LMXBs, although there is a growing population of transient HMXBs (McClintock et al. 2013; also for a review, see Remillard \& McClintock 
2006). 

The driving force of outbursts in transient black holes was a matter of debate for a long time and still is due to a lack of clear understanding. However, it is 
well known that viscosity plays the most important role in erupting an outburst (Ebisawa et al. 1996). When there is an enhancement of viscosity at the 
outer edge, observations of the outburst evolution suggest that matter approaches the compact object and starts an outburst (Chakrabarti 1996). 
The interplay between the thermal and tidal 
instabilities was put forward (Ichikawa et al. 1994) that explained the cause of an outburst and the variation of flux or counts in it. When there is an increase 
in viscosity, it sets in thermal instability in the disk. According to Chakrabarti et al. (2019), when the matter comes from the companion, it first starts to 
accumulate at a faraway location from the compact object hole due to a lack of viscosity. As more matter starts to accumulate, it increases the temperature 
and gradually, instability sets in. This increases the viscosity. As more matter gets accumulated, viscosity increases more, and as it crosses a critical level, 
matter starts to come towards the black hole and starts an outburst. The accumulation location is known as the pile-up radius. Using this concept of pile-up radius, 
we were able to explain the outbursting nature of a few black hole candidates, e.g., H 1743-322 (Chakrabarti et al. 2019), GX 339-4 (Bhowmick et al. 2021), the 
peculiar source 4U 1630-472 (Chatterjee et al. 2022). According to its nature, the BH outbursts are generally divided into two types (Debnath et al. 2010): the fast 
rise slow decay (FRSD) and slow rise slow decay (SRSD). Based on the rebrightening nature, Zhang et al. (2019) classified outbursts into different classes, such as 
glitch, reflare, multipeak, mini-outburst, or a new-outburst. 

The spectrum of a black hole binary comprises the combination of the soft multi-color thermal black body and hard non-thermal power-law components. The soft 
component is modeled after the radiation of the standard Keplerian disk (Shakura \& Sunyaev 1973, hereafter SS73), while one of the plausible explanations for the hard 
component comes from physics at play in the Comptonizing region, also known as the `Compton Cloud', which is the repository of hot electrons (Sunyaev \& Titarchuk 
1980, 1985). Many models, e.g., Bondi flow (Bondi 1952), standard-disk model (Shakura \& Sunyaev 1973), thick disk model (Paczynski \& Witta 1980), etc., came over 
the years to explain the composite spectrum of a stellar-mass black hole. Although they were quite successful in explaining so, moreover they all assumed some special 
conditions. A more general solution came in 1995 as Chakrabarti and his collaborators came up with the two-component advective flow (TCAF, Chakrabarti \& Titarchuk 
1995) solution. According to this model, the matter that comes from the companion star has two types of distribution of angular momentum, hence the name of the model. 
They are Keplerian and sub-Keplerian. The Keplerian component, being highly viscous, forms an accretion disk that resides on the equatorial plane. As viscosity 
increases, it starts to come towards the black hole. The sub-Keplerian one has less viscosity, and it radially falls in a free-fall timescale encapsulating the 
Keplerian disk. This optically thin sub-Keplerian component forms a shock front where matter is stopped due to the counterbalance between the pressure due to the 
gravitational and centrifugal forces. This shock front is also the boundary layer, known as the CENtrifugal pressure-supported BOundary Layer or CENBOL. This is the 
`Compton' cloud region in this model which is the repository of hot electrons. The Keplerian component gives the explanation of the soft multi-color black body 
component as the disk produces blackbody photons. The CENBOL acts as the Comptonizing region, which up-scatters some of the intercepted soft disk photons and releases 
them as the hard power-law photons.

During an outburst, a transient black hole goes through different states, having different spectral natures (Remillard \& McClintock 2006). Generally, we observe 
four different spectral states of a BH during an outburst. They are hard state (HS), hard intermediate state (HIMS), soft intermediate state (SIMS), and soft state 
(SS) (e.g., Nandi et al. 2012 and references therein). Sometimes in an active phase of the outburst during the declining state, very low counts are observed, and 
this is also known as the low hard state (LHS). Also, some sources show a very high count rate during the soft state phase, indicating the presence of a high soft 
state (HSS). During some outbursts, a source goes through all the above four generally observed spectral states and is known as type-I outbursts (Debnath et al. 2017). 
In the case of some other outbursts, sources only go through the harder spectral states and miss the softer ones. They are known as type-II or `failed' ones 
(Tetarenko et al. 2016; Debnath et al. 2017). The contribution from the soft black body and hard power-law radiations designate the nature of spectral states in 
outbursts. In the case of the hard state, the supply from the Keplerian disk photons remains low, and as a result, the CENBOL is not very much cooled due to up-scattering 
those photons. So, the temperature of the CENBOL remains high with low disk and halo accretion rates. When the supply of soft photons increases, the cooling of the 
Compton cloud increases, and gradually the source reaches a soft state. For a more detailed description of how the rates change in the case of various spectral states, 
see Chakrabarti \& Chattopadhyay (2000), Debnath et al. (2015). 

Both black holes and neutron stars show some common X-ray spectral features under certain conditions. However, previously, it was thought there was some 
special feature, the high-soft state spectrum, that plays as the distinction between them. In the high soft state of black holes, some sources showed high-energy 
power-law tails. Attempts were made to explain this feature within the framework of thermal Comptonization, where it was assumed that the disk photons were thermally 
up-scattered by the Compton cloud that is very close to the cold disk (Borozdin et al. 1999). However, this would have resulted in a power-law index of $\Gamma \sim 
1.5$, which is analogous to the hard state scenario. Thus it came to light that Comptonization due to the hot electrons in the `Compton cloud' is not the only 
Comptonization process that takes place in case of black hole outbursts. In the case of the high soft state, when the disk reaches very close to the compact black 
hole, another type of Comptonization may take place, which is due to the bulk motion of the cold matter in the Keplerian disk. In the case of the high soft state, when 
matter reaches inside the inner sonic point, the flow becomes convergent as it radially falls in a free-fall timescale. Very close to the black hole, as the bulk 
of the matter moves with relativistic speed, the relativistic cold electrons transport their momentum to the photons and lose energy in the process (Borozdin et al. 
1999). This is known as bulk motion Comptonization or BMC. This produces the long power-law tail in the high soft state spectrum of a black hole binary. The concept 
of BMC was first introduced by Blandford \& Payne (1981a,b); Payne \& Blandford (1981). Psaltis \& Lamb (1997) added relativistic corrections to the model transfer 
equations, which was later solved by Psaltis (2001) for flat space-time. Studies suggested that this BMC is observed only for high soft states for which the photon 
index of power-law $\Gamma \geq 3$ (Psaltis 2001). Chakrabarti \& Titarchuk (1995) and Ebisawa et al. (1996) suggested that the presence of high-energy power-law 
tails in the soft state spectrum is a distinctive signature of black hole binaries. However, later, it was found that neutron stars also show bulk motion 
Comptonization in HSS (Niedzwiecki \& Zdziarski (2006)). Previously, it was thought that there is a need for a horizon for the bulk Comptonization process. In case 
of neutron stars, matter from the innermost part of the disk to the neutron star falls with relativistic speed. Seed photons are Comptonized by this bulk motion at 
the innermost part of the neutron star surface, also known as the transition layer (Farinelli et al. 2008). This produces a power-law like tail that extends up to 
high energies. Zdziarski (2000), Zdziarski et al. (2001), McConnell et al. (2002), and Zdziarski \& Gierlinski (2004) suggested that the hard power-law tail in the 
soft state extends up to very high energies of hundreds of keV, even sometimes to MeV.  

The Galactic transient LMXB MAXI J0637-430 showed an outburst in 2019--20. The gas slit camera (GSC) on board the monitor of all-sky X-ray image (MAXI) discovered 
this source on November 02, 2019, during the onset of the outburst (Negoro et al. 2019). After its detection, this source was observed by many other satellites,
e.g., AstroSat, NuSTAR, NICER, Swift. The source is located at an R.A., Dec. = $99.09828^\circ, -42.8678^\circ$
(Kennea et al. 2019). According to Ma et al. (2022), this source belongs to the thick-disk population of the Milky Way BHs. This outburst continued for almost 6 
months. The earliest studies on this source suggested that this is in an LMXRB system (Strader et al. 2019) with a black hole as the primary compact object (Tomsick 
et al. 2019). After its discovery in X-ray, it was also monitored and observed in radio (Russell et al. 2019), optical (Li \& Kong 2019), and infrared (Murata et 
al. 2019) wavebands. Several authors reported the X-ray properties of the source using various archival public data (Thomas et al. 2019; Tomsick et al. 2019; 
Remillard et al. 2020; Jana et al. 2021; Lazar et al. 2021). Using NICER data, Jana et al. (2021) reported the complete spectral evolution of this outburst using 
phenomenological models. This source has gone through all the four defined spectral states in the following manner LHS $\rightarrow$ HIMS $\rightarrow$ SIMS 
$\rightarrow$ HSS $\rightarrow$ HIMS $\rightarrow$ SIMS $\rightarrow$ HSS. They also reported on the timing properties. Weak variability was observed during the 
SIMS and HSS, while the HS and HIMS showed strong variability (Jana et al. 2021). Lazar et al. (2021) analyzed the source with the combined Swift/XRT and NuSTAR/XTI 
data. They reported the presence of a reflection component in the soft state. The mass of this source is in the range of $5-12~M_\odot$, and it is situated at a 
distance of about 7kpc (Jana et al. 2021). It is located at a Galocentric distance of $\sim 13.2 \pm 1.8$ kpc (Soria et al. 2022), making it the farthest 
discovered Galactic black hole from the center. This is also at the height of $\sim 3.1 \pm 0.8$ kpc from the Galactic plane (Soria et al. 2022). They reported a 
distance of $\sim 8.7 \pm 2.3$ kpc with the BH mass of $\sim 5.1 \pm 1.6 ~M_\odot$ with a spin of $a \leq 0.25$ and a donor star of $M_2 \sim 0.25 \pm 0.07 ~M_\odot$. 
This source also happens to be the shortest-period black hole binary with an orbital period $P_{orb} \sim 2.2^{+0.8}_{-0.6}$ hr (Soria et al. 2022). 

Here in this work, we have analyzed simultaneous NICER+NuSTAR data in the broadband range of $0.7-70$ keV during the 2019--20 outburst of MAXI J0637-430. We have 
adopted both the phenomenological and physical approaches for this analysis. In \S2, we describe the data selection, data reduction, and data analysis process. In \S3, 
we show the results of our analysis. In \S4, we elaborate and discuss the results. Finally, in \S5, we make a summary of our work and draw conclusions from it.

\section{Observation, Data Selection, Reduction, and Analysis}

This source was observed and monitored by many X-ray satellites. For this work, we have used only NICER, Swift, and NuSTAR satellite data. We discuss data selection, 
data reduction, and data analysis in the next three subsections, respectively.

\subsection{Data Selection}

This source was monitored by NICER and XRT on a regular basis from the start to the end of the outburst between November 2019 to April 2020. There are more than 90 
observations in the case of both the NICER and XRT instruments. However, evolution was already studied for this source. Our aim was to study the broadband property, 
especially the high soft state property of the outburst. NuSTAR also observed this outburst in 8 epochs. For our broadband analysis, we needed simultaneous 
NICER--NuSTAR and/or XRT--NuSTAR data. We found that on 4 and 2 epochs, there were simultaneous observations from NICER--NuSTAR (MJD $\sim$ 58792, 58800, 58879, and 
58889) and XRT--NuSTAR (MJD $\sim$ 58812, and 58866) respectively. For this selection, we have chosen NICER observations 2200950103, 2200950108, 2200950163, \& 
2200950173, XRT observations 00012172018, \& 00012172066, and NuSTAR observations 80502324002, 80502324004, 80502324006, 80502324008, 80502324010, \& 80502324012 
respectively. There were a total of 8 NuSTAR observations. The other 2 NuSTAR observations were not taken into this analysis, as there were no simultaneous 
data below 3 keV for broadband analysis. In Table 1, we give the full data observation list.

\begin{table*}
\small
 \addtolength{\tabcolsep}{-1.5pt}
 \centering
 \caption{List of Data used}
 \label{tab:table1}
 \begin{tabular}{cccccccc}
 \hline
 UT$^{[1]}$  &    Day$^{[1]}$  &    NICER    &    NICER   &    XRT        &    XRT   &    NuSTAR      &   NuSTAR    \\
 Date        &    MJD          &   Obs. Id.  &   Exp. (s) &   Obs. Id.    & Exp. (s) &   Obs. Id.     &  Exp. (s)   \\
  (1)        &     (2)         &     (3)     &     (4)    &    (5)        &    (6)   &      (7)       &     (8)     \\
 \hline
 \hline 
 2019-11-05  &    58792.5      & 2200950103  &      570   &      -        &    -     &  80502324002   &   36799     \\ 
 2019-11-13  &    58800.6      & 2200950108  &     1821   &      -        &    -     &  80502324004   &   67738     \\ 
 2019-11-25  &    58812.0      &     -       &      -     &  00012172018  &  1669.6  &  80502324006   &   48626     \\ 
 \hline 
 2020-01-18  &    58866.0      &     -       &      -     &  00012172066  &   674.6  &  80502324008   &   46630     \\ 
 \hline
 2020-01-31  &    58879.6      & 2200950163  &     2876   &      -        &    -     &  80502324010   &  110779     \\ 
 2020-02-10  &    58889.6      & 2200950173  &     1281   &      -        &    -     &  80502324012   &   50233     \\ 
\hline
\hline 
 \end{tabular}

\vspace{0.2cm}
 \noindent{
         \leftline{$^{[1]}$ Columns 1 and 2 represent the universal time and MJD of the data used.}
         \leftline{Columns 3, 5, and 7 represent observation ids from NICER, XRT, and NuSTAR satellites/instruments, respectively.}
         \leftline{Columns 4, 6, and 8 represent exposure times of observation ids from NICER, XRT, and NuSTAR satellites/instruments respectively.}
	 \leftline{The horizontal lines in the table signify different states during different epochs of this outburst. The first 3 epochs are from HSS,}
         \leftline{4th is from HIMS, and last 2 are from declining HS.}
          }
\end{table*}

\subsection{Data Reduction}

\subsubsection{NICER}

We take data from the X-ray timing instrument (XTI) onboard the neutron star interior composition explorer (NICER; Gendreau et al. 2012; Prigozhin et al. 2012) at 
the international space station (ISS). This instrument covers an energy range of $0.2-10$ keV with an energy resolution of $\sim 85$ eV at 1 keV. It has a very 
high timing resolution of $\sim 100$ ns. It provides a very high signal-to-noise (S/N) ratio. With the calibration file (version 20200722; package
{\fontfamily{pcr}\selectfont goodfiles\_nicer\_xti\_20200722.tar.gz}), we first run the {\fontfamily{pcr}\selectfont nicerl2} task to reprocess clean event files from 
uncleaned archive data. After this step, they are barycentre corrected. Using the barycentre corrected cleaned event files, we extract the spectrum and light curves 
using {\fontfamily{pcr}\selectfont XSELECT} task of {\fontfamily{pcr}\selectfont FTOOLS} package of {\fontfamily{pcr}\selectfont HEASoft (version 6.29)} 
software. Using the {\fontfamily{pcr}\selectfont nibackgen3C50} (Remillard et al. 2022) task, we produce the background files for our analysis. At last, we 
grouped the data using the {\fontfamily{pcr}\selectfont GRPPHA} task with a minimum of 20 counts per spectral bin.

\subsubsection{Swift/XRT}

We used windowed timing (WT) data for Swift's X-ray telescope (XRT) instrument. First, we run the {\fontfamily{pcr}\selectfont xrtpipeline} command to produce cleaned
event files from the stage-I uncleaned data. From each cleaned event file, using {\fontfamily{pcr}\selectfont XSELECT} task, we choose source and background region 
files in {\fontfamily{pcr}\selectfont ds9}. A circular radius of $30$ arcsec was used for selecting both the source and background regions. Then, using the source 
and background region files, we extract the light curve and spectrum files. The detailed data reduction method is mentioned in the 
\href{https://www.swift.ac.uk/analysis/xrt/}{XRT data analysis thread} page. At last, we grouped the data using {\fontfamily{pcr}\selectfont GRPPHA} task 
with a group min of 20 counts/bin for the windowed timing (WT) mode. We use the XRT calibration data file (version 20190910; package 
{\fontfamily{pcr}\selectfont goodfiles\_swift\_xrt\_20190910.tar.Z}) for the full data extraction process. For WT mode data, spectrum does not get affected 
by pile-up for intensities below 100 counts/sec. For both of our XRT observations, the intensity was below 100 counts/sec. Thus we did not take pile-up correction
into consideration.

\subsubsection{NuSTAR}

We use data from the focal plane module (FPM) instrument onboard the NuSTAR satellite. We use the analysis software {\fontfamily{pcr}\selectfont NuSTARDAS} (version 
1.4.1) and calibration files (version 20200626; package {\fontfamily{pcr}\selectfont goodfiles\_nustar\_fpm\_20200626.tar.gz}) to perform the data reduction 
process. The {\fontfamily{pcr}\selectfont nupipeline} task was first run to produce stage-II cleaned event files using the stage-I data providing input, steminputs, 
and output directories. Using the stage-II cleaned event files, we make use of the {\fontfamily{pcr}\selectfont XSELECT} task to locate the source and background 
region using {\fontfamily{pcr}\selectfont ds9} and save those region files. We use a circular region (as no pile-up was detected) of 80 arcsec radius for both the 
source and background regions. Using those region files, we then run the {\fontfamily{pcr}\selectfont nuproducts} command to extract stage-III spectrum and light 
curve files. Then we use the {\fontfamily{pcr}\selectfont GRPPHA} task for grouping the data with a minimum count rate of 30 counts/s per spectral bin.

\subsection{Data Analysis}

We mainly perform spectral analysis of the BHC MAXI J0637-430 during its 2019--20 outburst. 
The daily average MAXI/GSC light curve is downloaded from the public archive. Using simultaneous 
NICER--NuSTAR and XRT--NuSTAR data, we have fitted a total of 6 observed spectra in the broadband range of $0.7-70$ keV in {\fontfamily{pcr}\selectfont XSPEC}. 
For the combined spectral analysis in a broad range, we have considered NICER and Swift/XRT in the $0.7-8$ keV and NuSTAR in the $7-70$ keV energy bands, respectively. 
The NICER data suffers from residuals in lower energies due to absorption edge features of Si in $1.7-2.1$ keV, Au in $2.2-2.3$ keV, etc., typical for Si-based 
detectors, which could be due to instrumental error or error in the response matrix. These are more prominent in the case of bright sources (e.g., Swift J0243.6+6124;
Jaiswal et al. 2019). This could result in some residuals in the lower energy part of the spectra. Since there is much noise (low signal-to-noise ratio) in the data 
above 70 keV for NuSTAR data, we have taken the NuSTAR spectrum up to only 70 keV. We first considered the phenomenological approach and then took the physical model 
approach. The phenomenological models give a rough idea about the properties
of the system without giving much physical insight into the system. Thus we reanalyzed using the physical models, which give the idea of the physical properties of the 
source and also tried to confirm the findings from the phenomenological approach. To account for the interstellar absorption, we make use of the 
{\fontfamily{pcr}\selectfont TBABS} model considering {\fontfamily{pcr}\selectfont wilms} abundance (Wilms et al. 2000). We have also spectrally fitted using
the absorption model {\fontfamily{pcr}\selectfont tbnew\_feo} of the {\fontfamily{pcr}\selectfont absmodel}
\href{https://pulsar.sternwarte.uni-erlangen.de/wilms/research/tbabs/}{package}. This is an updated version of the {\fontfamily{pcr}\selectfont TBABS} model. Along 
with having the hydrogen column density ($n_H$), this model also considers the variations in the abundances of O and Fe relative to Solar abundance. This is reported 
in the Appendix section. In the case of the high soft state, we first 
fit the data using the {\fontfamily{pcr}\selectfont tbabs(diskbb+power-law)} model using only the NICER data in the $0.7-10$ keV energy band. The motivation 
behind using this combination of models is due to the fact that the composite spectrum of a black hole comes primarily from the contribution of the disk and the 
coronal contribution. Thus we wanted to check the primary emission using this combination of models, where the {\fontfamily{pcr}\selectfont diskbb} component 
models the spectrum for the contribution from the blackbody radiation from the disk, whereas the {\fontfamily{pcr}\selectfont power-law} component models the 
contribution from the Comptonization component. However, when we considered NICER+NuSTAR in the broadband range ($0.7-70$ keV), we modified and applied the 
{\fontfamily{pcr}\selectfont constant*tbabs(diskbb+broken power-law)} model, as the only {\fontfamily{pcr}\selectfont power-law} model was not able to fit the data 
for an acceptable $\chi^2/DOF$. The {\fontfamily{pcr}\selectfont broken power-law} is generally used when there are different slopes in the spectrum instead 
of a single slope with an energy break ($E_{break}$). This could be due to the contribution from multiple radiation processes instead of only inverse-Comptonization. 
Then, for the physical approach, we applied the {\fontfamily{pcr}\selectfont tbabs(TCAF+power-law)} model for NICER data in the $0.7-10$ keV energy band. The 
significance of the TCAF model is described in detail in the introduction section. However, in this case also, we had to replace the power-law 
model with the broken power-law model for the same reason as the phenomenological approach. For NICER+NuSTAR in the $0.7-70$ keV energy band, the model 
tested, is {\fontfamily{pcr}\selectfont constant*tbabs(TCAF + broken power-law)}. For combined XRT--NuSTAR data also, in the high soft state, we fitted first with 
the {\fontfamily{pcr}\selectfont constant*tbabs(diskbb+broken power-law)} model and then refitted with the {\fontfamily{pcr}\selectfont constant*tbabs(TCAF + broken 
power-law)} model. For these observations, we also spectrally fitted with the {\fontfamily{pcr}\selectfont constant*tbabs(diskbb+power-law+bmc)} and 
{\fontfamily{pcr}\selectfont constant*tbabs(TCAF+power-law+bmc)} models. The {\fontfamily{pcr}\selectfont bmc} component is used to model the spectrum to 
check if there is a contribution from the bulk motion Comptonization process. However, for those observations, which are outside the high soft state phase, we used 
{\fontfamily{pcr}\selectfont constant*tbabs(diskbb+power-law)} and {\fontfamily{pcr}\selectfont constant*tbabs*TCAF} models for both the NICER--NUSTAR and XRT--NuSTAR 
data. As per previous reports, for our analysis epochs, the source was in a high soft state on MJDs 58792.5 (2019 May 11), 58800.6 (2019 May 13), and 58812 (2019 
May 25) respectively. We discuss our analysis results in more detail in the next section.

\section{Results}

We have used MAXI/GSC data to study the outburst profile in different energy bands and the evolution of the hardness nature of MAXI J0637-430 during its 2019--20 
outburst. To study the accretion flow properties of the source during the outburst we have analyzed its broadband spectra using archival NICER, XRT, and NuSTAR data. 
For our analysis, we first fitted the combined NICER+NuSTAR and XRT+NuSTAR data in $0.7-70$ keV energy band with both phenomenological and physical models. From 
those fitting, we have found some significant properties about the spectral nature of the source, which we discuss here.

\subsection{Outburst Profile and Hardness Ratio}

\begin{figure*}
\vspace{0.8cm}
  \centering
  \textbf{Evolution of Light Curve and Hardness Ratio}\par\medskip
    \includegraphics[width=12cm]{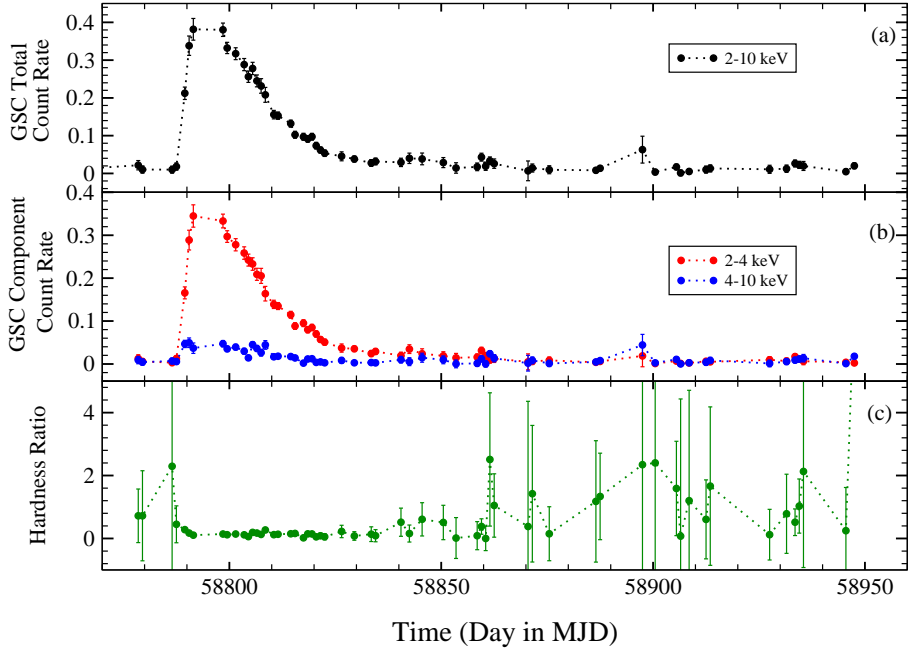}
    \caption{Variation of daily average MAXI/GSC (a) $2-10$ keV total count rate, (b) $2-4$ keV \& $4-10$ keV count rates, and (c) hardness ratio (HR) of 
	     $4-10$ keV to $2-4$ keV GSC count rates throughout the entire outburst.}
\end{figure*}

In Figure 1, we show the outburst light curve profile as the variation of MAXI/GSC count rates in various energy bands. Panel (a) shows the total count rate in the 
$2-10$ keV range, while in panel (b), we show the variations of the $2-4$ keV (red) and $4-10$ keV (blue) count rates, respectively. In panel (c), we show 
the variation of the hardness ratio (HR), which is the ratio of count rates of $4-10$ keV to $2-4$ keV energy bands. From the outburst profile, we observe that the 
nature of the outburst is a fast-rise slow decay (FRSD) type. At the start, it increased very rapidly within a few days and then very slowly decayed into quiescence. 
During the phase from $\sim$ 58787 to 58860, HR was low ($\leq 0.5$) when there is the dominance of soft band ($2-4$ keV) counts over the hard band ($4-10$ keV) 
counts in the GSC light curve. This phase was reported to have the presence of a high soft state (HSS) (Jana et al. 2021). Since several authors have studied this source, 
its spectral evolution has already been designated. Thus, we do not aim to spectrally classify this outburst for the entire phase. Our goal is to show from 
broadband spectral fitting that in HSS, there could be the presence of another component, the bulk motion Comptonization, to be specific.

\subsection{Spectral Properties}

Studying the spectra of a BHC during an outburst gives a clear idea about the spectral nature and evolution throughout that outburst. With the available NICER 
observations, Jana et al. (2021) reported that this source has gone through all the defined spectral states of a BHC in the following manner: SIMS $\rightarrow$ 
HSS $\rightarrow$ HIMS $\rightarrow$ HS during the entire phase of the outburst. According to their analysis, from the starting NICER observation day to 2019 
November 06 ($\sim$ MJD 58793), this was in SIMS in the rising phase. From that day to 2020 January 10 ($\sim$ MJD 58858), the source was in HSS, and the source was 
already in the declining phase. From 2020 January 10 (MJD 58858) to 2020 January 29 ($\sim$ MJD 58877), it was in HIMS and then it was in the declining hard state 
(HS). We have taken 4 simultaneous NICER+NuSTAR observations and 2 XRT+NuSTAR observations for our broadband analysis of this source during six epochs. Out of these 
total 6 observations, 2 NICER+NuSTAR and 1 XRT+NuSTAR observations are from the HSS state, 1 XRT+NuSTAR observation is from the HIMS state, and rest 2 NICER+NuSTAR 
observations are from the declining HS. This is listed in Table 1, and different states are divided using horizontal lines.

\begin{figure*}
\vskip 0.2cm
\centering
\textbf{Model Fitted NICER Spectra}\par\medskip
\vbox{
\includegraphics[width=8.5truecm,angle=0]{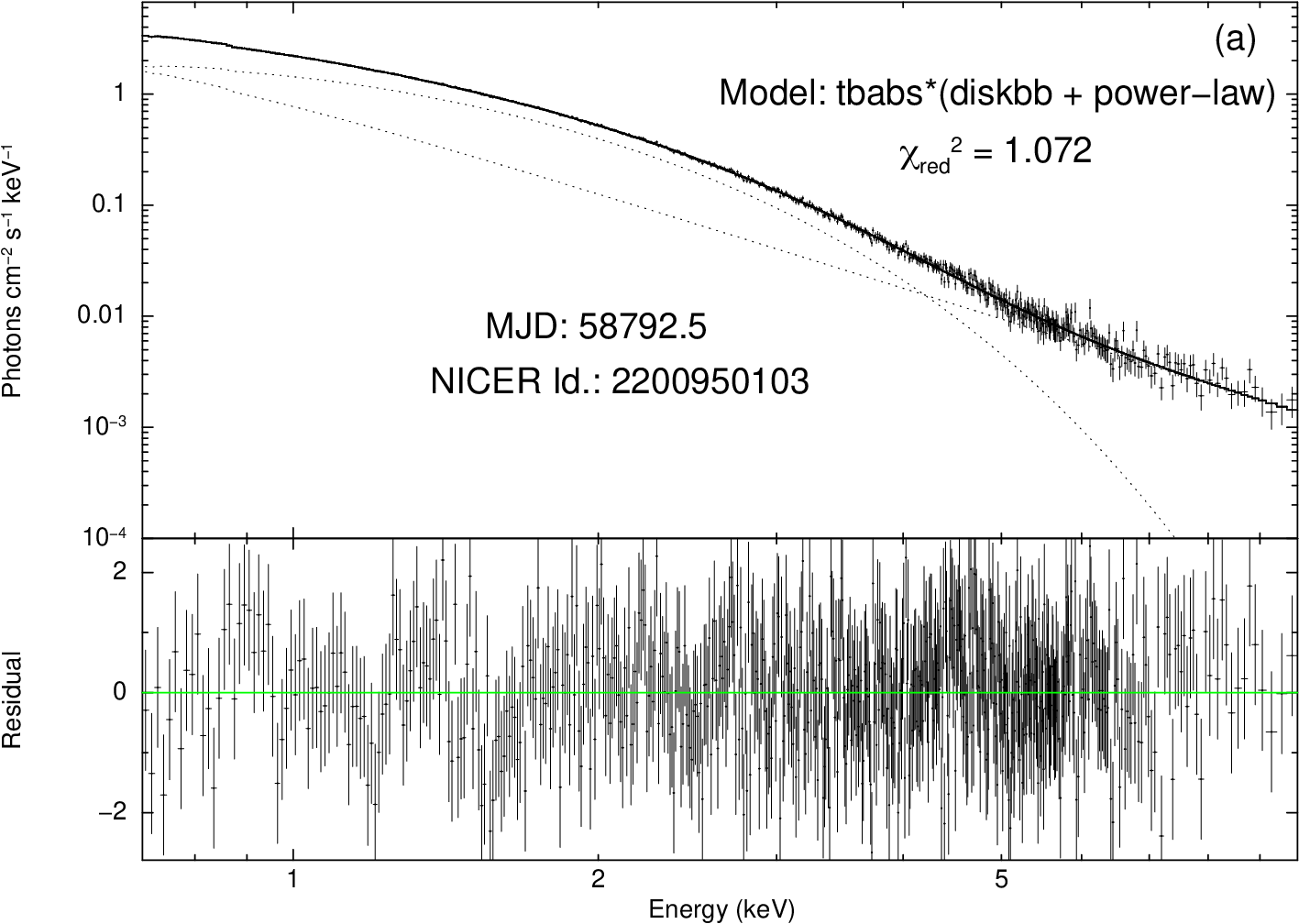}}
\vskip 0.1cm
\hskip 0.5cm
\vbox{
\includegraphics[width=8.5truecm,angle=0]{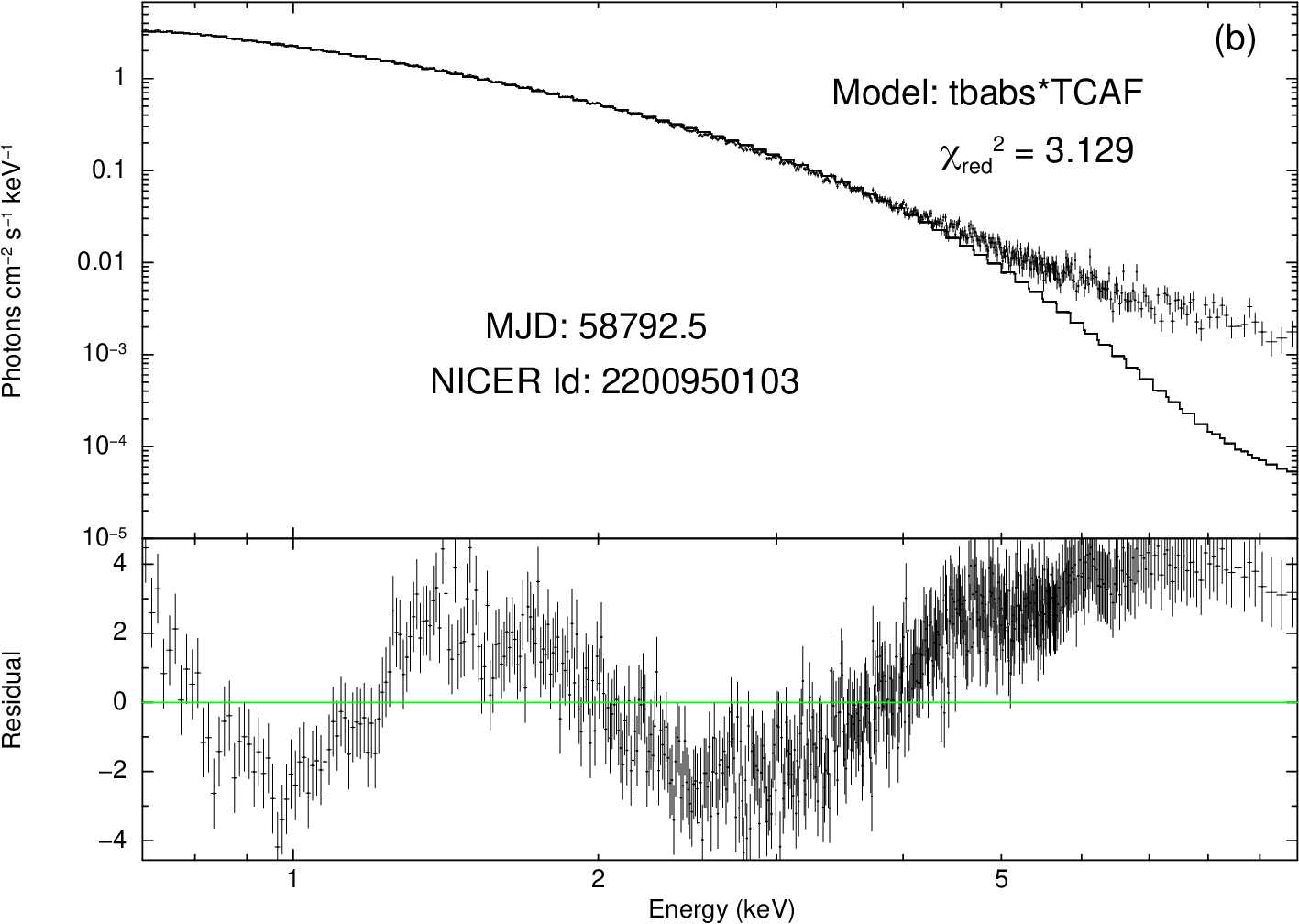}\hskip 0.5cm
\includegraphics[width=8.5truecm,angle=0]{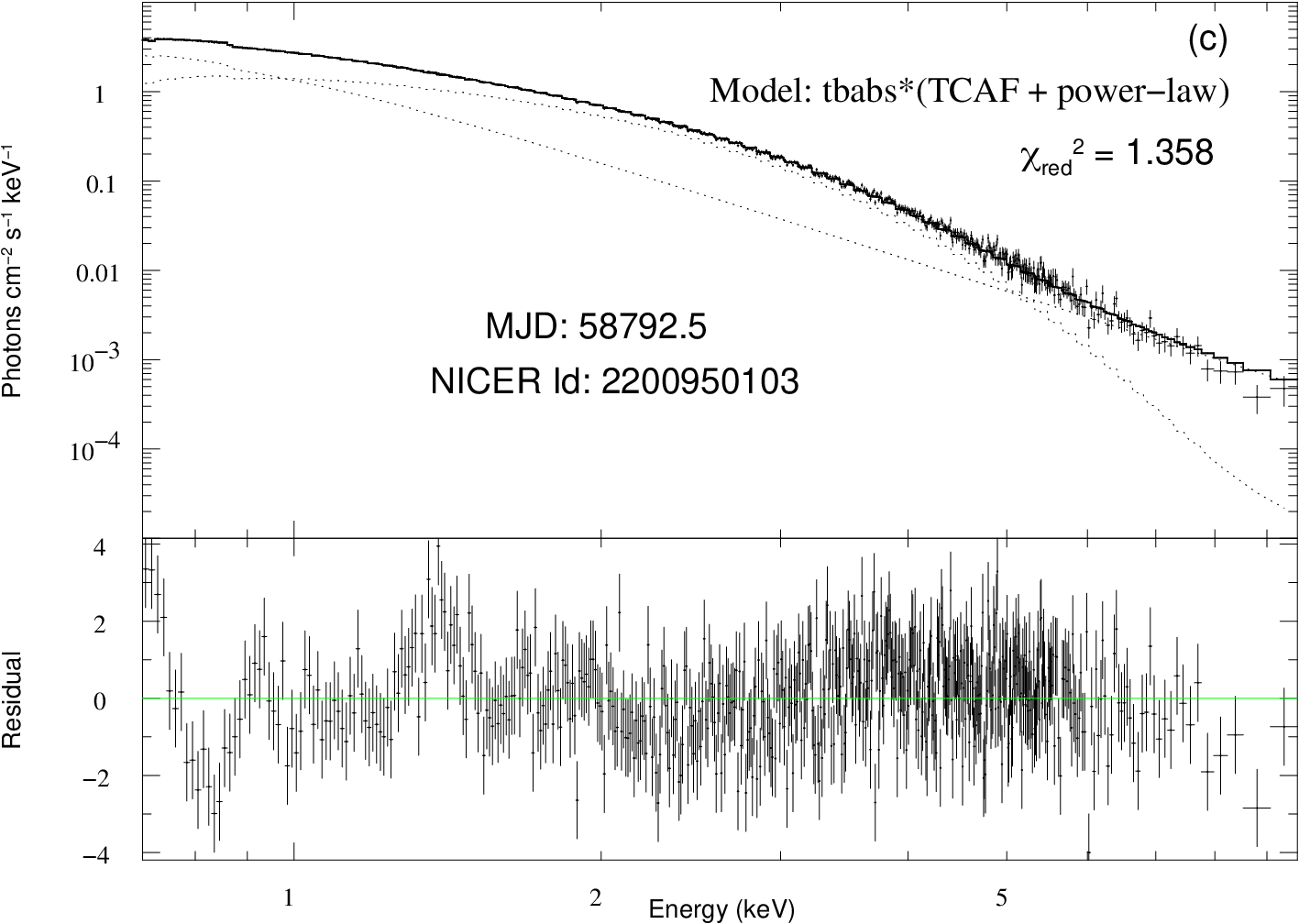}}
\caption{Model fitted unfolded NICER spectra using (a) {\fontfamily{pcr}\selectfont DBB+PL}, (b) {\fontfamily{pcr}\selectfont TCAF}, (c) {\fontfamily{pcr}\selectfont 
	 TCAF+PL} models in $0.7-10$ keV energy band. For all three cases, we use {\fontfamily{pcr}\selectfont tbabs} to account for the interstellar absorption.
	 All these spectra are shown from the NICER observation id 2200950103 on MJD 58792.5.}
\end{figure*}

We first fitted the NICER observation ID 2200950103 in the $0.7-10$ keV energy band using both the phenomenological {\fontfamily{pcr}\selectfont DBB+PL} and 
physical {\fontfamily{pcr}\selectfont TCAF} models. The {\fontfamily{pcr}\selectfont DBB+PL} model could fit the spectral data with acceptable best-fit statistics 
of $\chi^2/DOF = 1.072$. We found that a very high value of the photon index of power-law ($\Gamma = 3.22 \pm 0.12$) was required to spectrally fit the data within 
this acceptable range. For this fit, the inner-disk temperature $T_{in}$ was $0.615 \pm 0.001$ keV. These values support that the source was in HSS during this time. 
The fitted spectrum is shown in Figure 2a. However, when we fit the data with the {\fontfamily{pcr}\selectfont TCAF} model, we faced difficulty in achieving acceptable 
statistics for the fitting. Fitting with an absorbed {\fontfamily{pcr}\selectfont TCAF} model resulted in a $\chi^2/DOF = 3.129$, which shows a lot of 
residuals in the entire energy range (Fig. 2b). Using only the {\fontfamily{pcr}\selectfont TCAF} model, the best fit was not achieved. So, we added an additional 
{\fontfamily{pcr}\selectfont power-law} model with the existing {\fontfamily{pcr}\selectfont TCAF} model and found out that with the combined 
{\fontfamily{pcr}\selectfont TCAF+PL} fitting, a statistically acceptable $\chi^2/DOF = 1.358$ fit (Fig. 2c) was achieved. We think that residuals up to $\sim 2$ keV 
may have arisen due to the error in the response function for that particular observation. However, as we notice at the higher energy part, we can observe that the 
spectrum gives a hint of a change in the slope. For this observation, the {\fontfamily{pcr}\selectfont power-law} has an index of $\Gamma = 3.69 \pm 0.19$. For this 
fitting, we found that high values of both the disk and halo rates (${\dot m}_d = 9.69 \pm 0.07$, ${\dot m}_h = 3.10 \pm 0.03$) with clearly dominating disk over halo. 
The shock was weak with a very close boundary to the black hole at $X_s = 32.57 \pm 0.03 ~r_s$ and $R= 1.64 \pm 0.17$. This also supports the presence of HSS during 
this epoch. We checked all the NICER data during the HSS phase of the outburst. We took a total of 44 observations during the HSS period and fitted them with both 
the {\fontfamily{pcr}\selectfont DBB+PL}, {\fontfamily{pcr}\selectfont TCAF}, and {\fontfamily{pcr}\selectfont TCAF+PL} models. For all these observations, we observed 
that $\Gamma$ needed a high value $\sim 3-4$ (except only for 3 observation ids, where $\Gamma \sim 2.5-2.9$) to achieve statistically acceptable best-fit. For all 
these observations, only the {\fontfamily{pcr}\selectfont TCAF} model was unable to fit the data, and the best fit was achieved only by the combined 
{\fontfamily{pcr}\selectfont TCAF+PL} model fitting. For all the {\fontfamily{pcr}\selectfont TCAF+PL} model fitted observations, the shock was weak with $R = 
1.05-1.9$ and was at a distance of $X_s= 22-33 ~r_s$. Both the accretion rates gradually decreased and varied in the range as ${\dot m}_d = 9.7-1.3$ and 
${\dot m}_h= 3.8-0.9$. However, reporting the full evolution of the outburst is not the aim of this paper.

The need for an extra {\fontfamily{pcr}\selectfont power-law} model with the {\fontfamily{pcr}\selectfont TCAF} model and the change of spectral slope in 
soft X-ray band data of NICER and XRT motivated us to check the broadband spectra to see if any other component was present in the high soft state for this outburst. 
With the available NuSTAR data, we took 4 simultaneous NICER--NuSTAR and 2 simultaneous XRT-NuSTAR data and fitted in the $0.7-70$ keV energy band. Above $70$ keV, 
NuSTAR data have noise (low s/n ratio) for this source. We first fitted the broadband NICER--NuSTAR spectrum on 5 November 2019 (MJD 58792.5). Although the 
{\fontfamily{pcr}\selectfont DBB+PL} model with a high $\Gamma$ was able to fit the NICER data in the $0.7-10$ keV energy range, we found that for a higher energy 
band, this model did not return a very good fitting. For this fit, we got $\chi^2/DOF=1.484$. However, an excess over 40 keV was seen, signifying a change in spectral 
slope. Hence we replaced the {\fontfamily{pcr}\selectfont power-law} with the {\fontfamily{pcr}\selectfont broken power-law (bknpl)} model and fitted it with the 
{\fontfamily{pcr}\selectfont DBB+BKNPL} model. The fit improved significantly now with $\chi^2/DOF=1.174$. This is shown in Figure 3(a--b). 

For the broadband fitting with the physical model, the situation was similar. However, here we observed the need to add another component. Previously for the 
$0.7-10 ~keV$ energy band spectral fitting, the {\fontfamily{pcr}\selectfont TCAF+PL} model was returning acceptable statistics for fit with $\chi^2/DOF \sim 1$. 
However, when we fitted using the {\fontfamily{pcr}\selectfont TCAF+PL} model in the $0.7-70$ keV energy band, the fitting returned bad statistics with $\chi^2/DOF 
= 2.116$. Like the phenomenological case, we now modified and changed the {\fontfamily{pcr}\selectfont power-law} model with the {\fontfamily{pcr}\selectfont 
broken power-law} and the fit returned an acceptable fit-statistic with $\chi^2/DOF = 1.194$. These spectra are shown in Figure 4(a--b).

\begin{figure*}
\vskip 0.2cm
\centering
\textbf{Model Fitted NICER+NuSTAR Spectra}\par\medskip
\vbox{
\includegraphics[width=8.5truecm,angle=0]{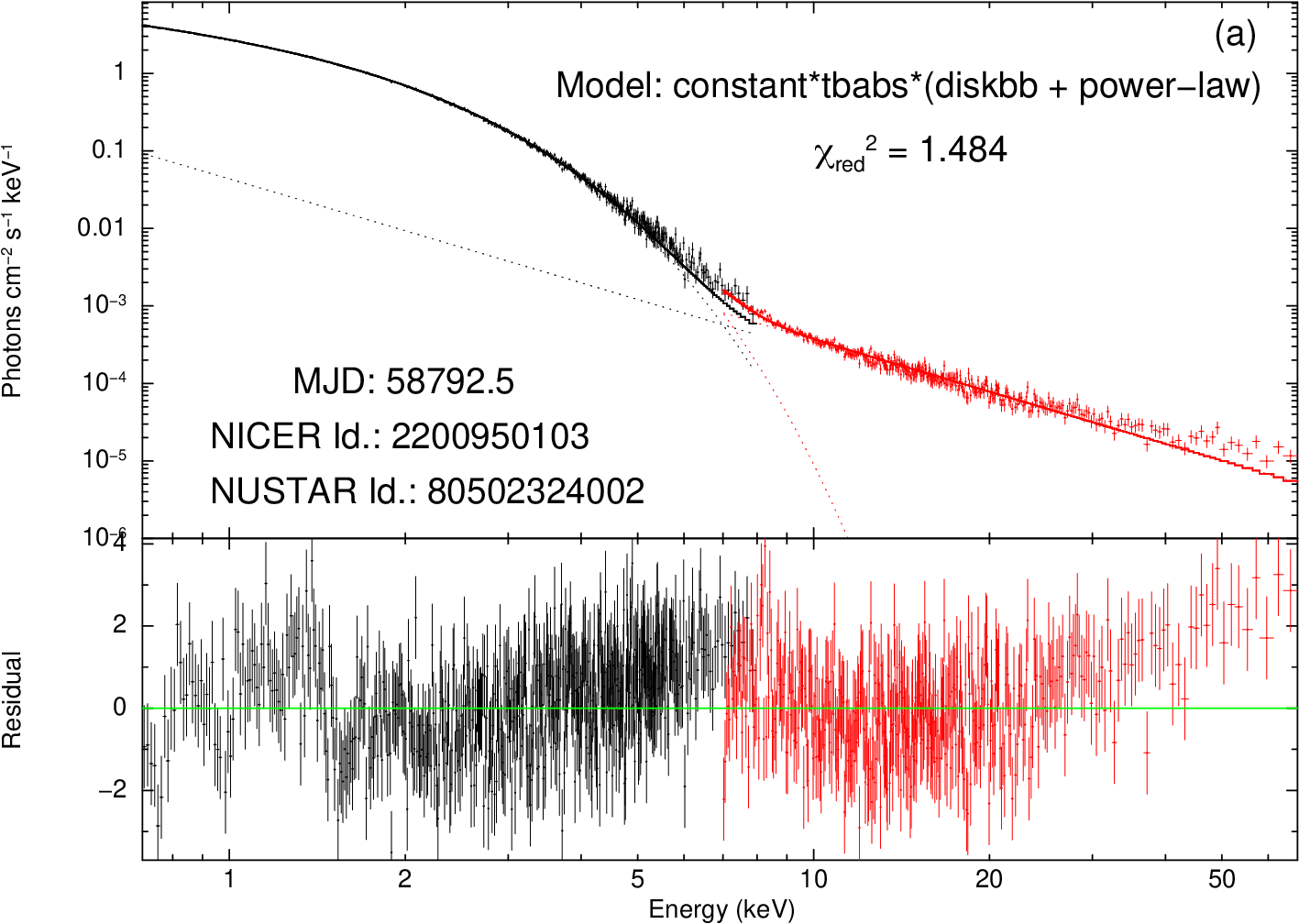}\hskip 0.5cm
\includegraphics[width=8.5truecm,angle=0]{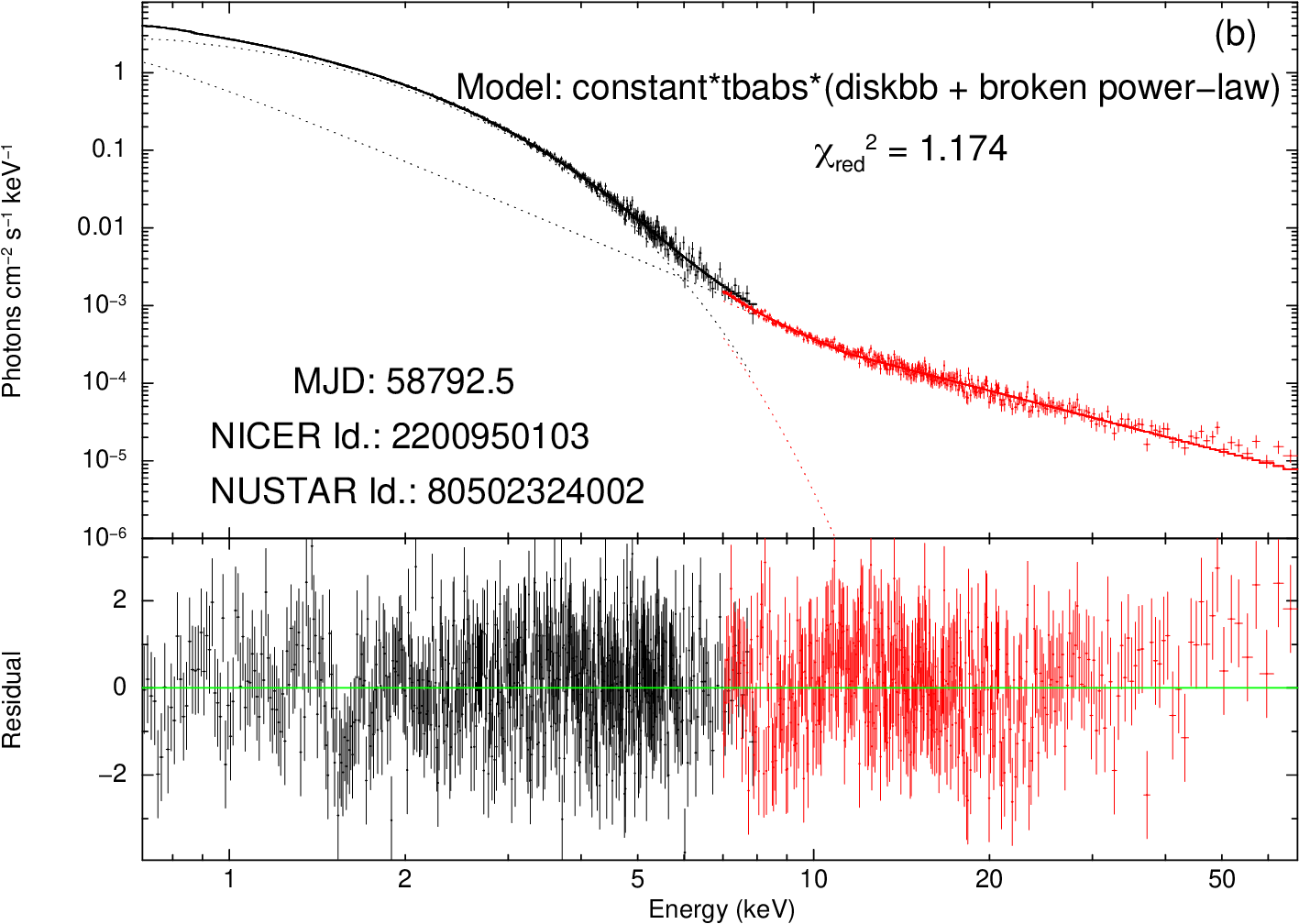}}
\caption{Model fitted unfolded NICER+NuSTAR spectra using (a) {\fontfamily{pcr}\selectfont DBB+PL}, and (b) {\fontfamily{pcr}\selectfont DBB+BKNPL} models in 
	 $0.7-70$ keV energy band. For all three cases, we use {\fontfamily{pcr}\selectfont tbabs} to account for the interstellar absorption. Both these 
	 spectra are shown from the combined NICER and NuSTAR observation ids 2200950103 and 80502324002 respectively, on MJD 58792.5.}
\end{figure*}

\begin{figure*}
\vskip 0.2cm
\centering
\textbf{Model Fitted NICER+NuSTAR Spectra}\par\medskip
\vbox{
\includegraphics[width=8.5truecm,angle=0]{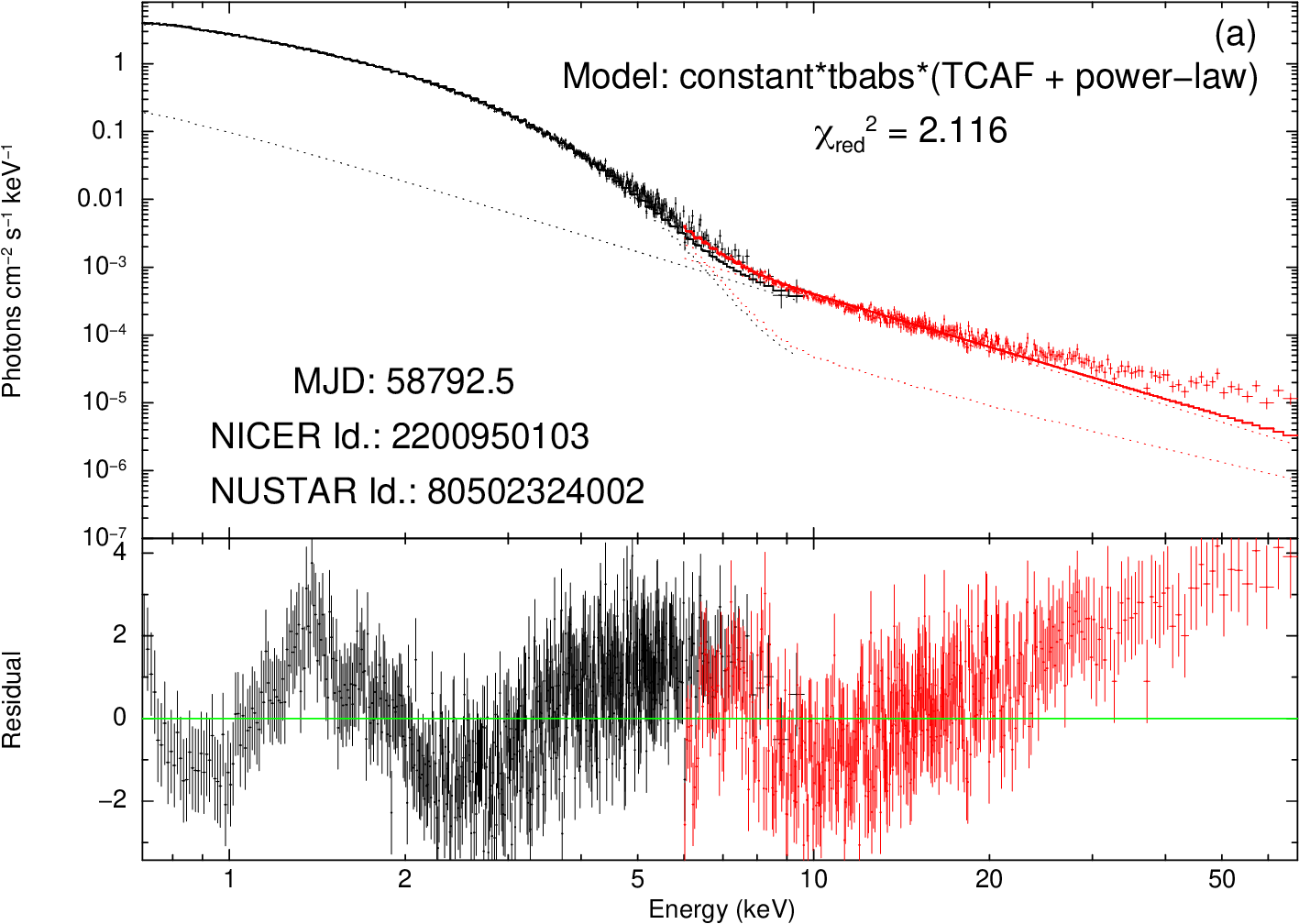}\hskip 0.5cm
\includegraphics[width=8.5truecm,angle=0]{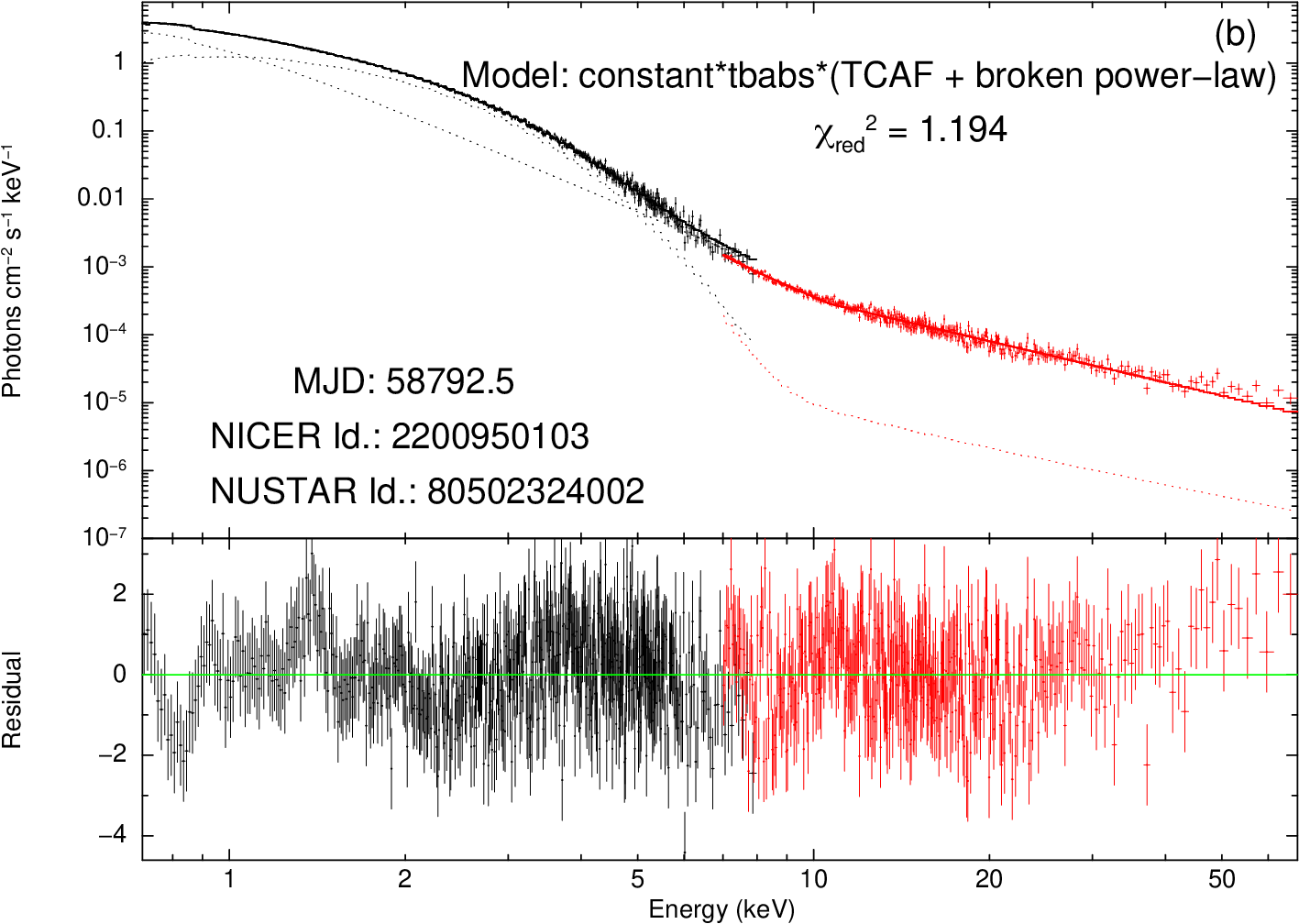}}
\caption{Model fitted unfolded NICER+NuSTAR spectra using (a) {\fontfamily{pcr}\selectfont TCAF+PL}, and (b) {\fontfamily{pcr}\selectfont TCAF+BKNPL} models in 
	 $0.7-70$ keV energy band. For all three cases, we use {\fontfamily{pcr}\selectfont tbabs} to account for the interstellar absorption. Both these 
         spectra are shown from the combined NICER and NuSTAR observation ids 2200950103 and 80502324002 respectively, on MJD 58792.5.}
\end{figure*}

We faced the same difficulty while fitting the other NICER--NuSTAR and XRT--NuSTAR observations of HSS on 2019 November 13 (MJD 58800.6) and 2019 November 25 
(MJD 58812.0), respectively. When we fitted with only the {\fontfamily{pcr}\selectfont DBB+PL} and {\fontfamily{pcr}\selectfont TCAF+PL} models, the fittings 
returned bad fitting statistics. However, modifying with the {\fontfamily{pcr}\selectfont broken power-law} model in place of the {\fontfamily{pcr}\selectfont 
power-law} improved the fitting significantly. On MJD 58800.6 and 58812, {\fontfamily{pcr}\selectfont DBB+PL} fitting resulted in $\chi^2/DOF=$ 5.61 and 2.31 
respectively, whereas {\fontfamily{pcr}\selectfont DBB+BKNPL} fitting improved the statistics with $\chi^2/DOF=$ 0.96 and 1.26 on those same MJDs. For 
the physical fitting scenario, the {\fontfamily{pcr}\selectfont TCAF+BKNPL} model returned $\chi^2/DOF=$ values as 1.01 and 1.15, respectively for MJD 
58800.6 and 58812. The best-fitted results are given in Table 2. 

Outside of this HSS phase, we also fitted 1 XRT--NuSTAR and 2 NICER--NuSTAR observations, as mentioned in Table 1. The XRT--NuSTAR data on MJD 58866 could be well 
fitted with only the {\fontfamily{pcr}\selectfont DBB+PL} and {\fontfamily{pcr}\selectfont TCAF} models that resulted in acceptable best-fits with $\chi^2/DOF=$ 
1.00 and 1.02, respectively. The NICER-NuSTAR observations on MJD 58879.6 and 58889.6 were first fitted with only the {\fontfamily{pcr}\selectfont power-law} model 
and returned $\chi^2/DOF$ values of 1.417 and 1.224 respectively. However, adding a disk black body component improved the fitting significantly for the 1st one 
and marginally for the 2nd one. For this {\fontfamily{pcr}\selectfont DBB+PL} fitting, the $\chi^2/DOF$ values improved to 1.121 and 1.161 respectively. For the 
physical model fitting scenario, only the TCAF model was able to fit all these 3 XRT-NuSTAR (1) and NICER-NuSTAR (2) observations data. For these 3 data, 
{\fontfamily{pcr}\selectfont TCAF} model fitting returned $\chi^2/DOF$ values as 1.17, 1.41, and 1.27 respectively. All the fit statistics are given in Table 3. 

\begin{figure*}
\vskip 0.2cm
\centering
\textbf{Model Fitted NICER+NuSTAR Spectra}\par\medskip
\vbox{
\includegraphics[width=8.5truecm,angle=0]{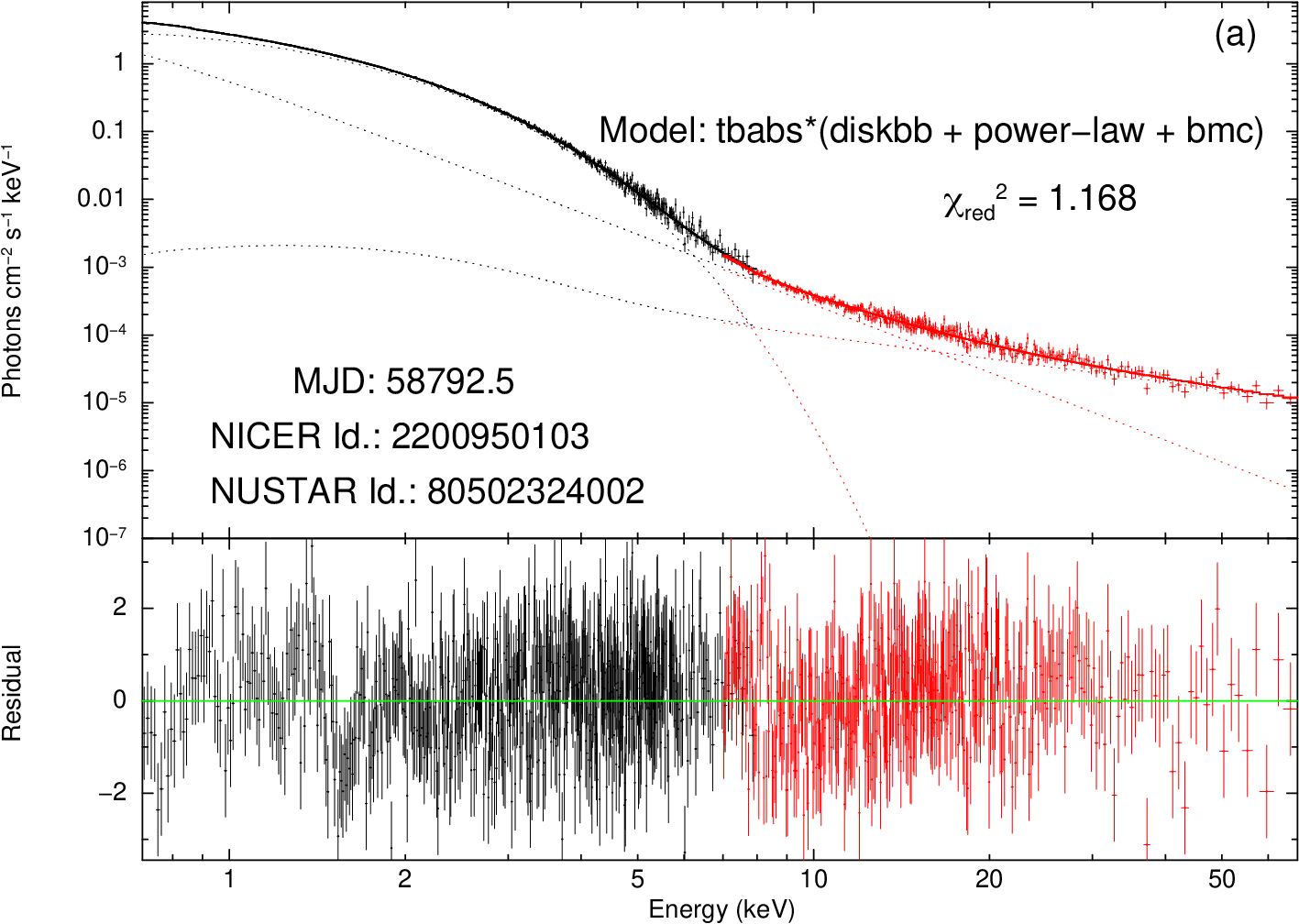}\hskip 0.5cm
\includegraphics[width=8.5truecm,angle=0]{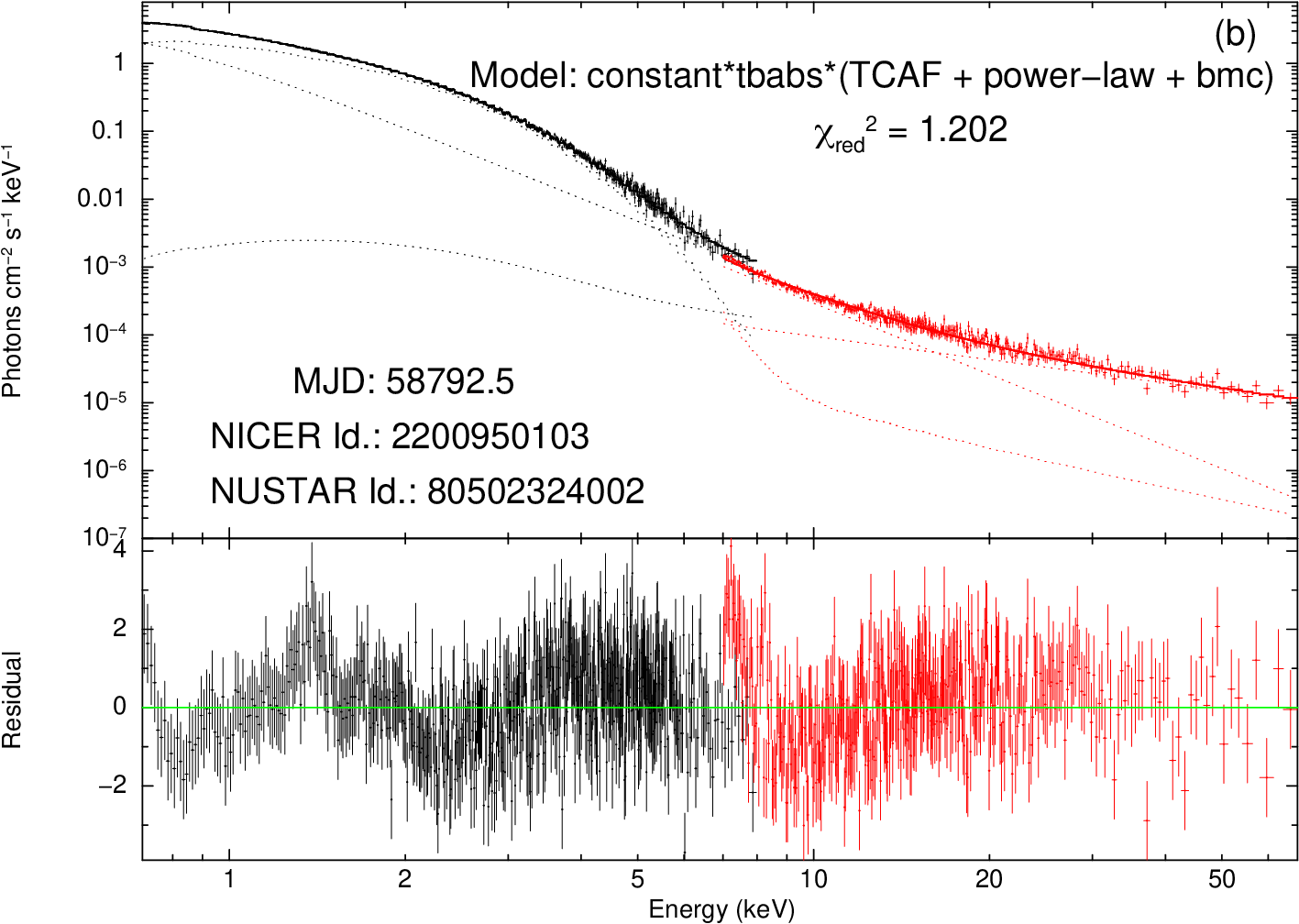}}
\caption{Model fitted unfolded NICER+NuSTAR spectra using (a) {\fontfamily{pcr}\selectfont DBB+PL+BMC}, and (b) {\fontfamily{pcr}\selectfont TCAF+PL+BMC} models in 
	 $0.7-70$ keV energy band. For all three cases, we use {\fontfamily{pcr}\selectfont tbabs} to account for the interstellar absorption. Both these 
         spectra are shown from the combined NICER and NuSTAR observation ids 2200950103 and 80502324002 respectively, on MJD 58792.5.}
\end{figure*}

Here we first report our result in the HSS from the best-fitted models using both the phenomenological {\fontfamily{pcr}\selectfont DBB+BKNPL} and physical 
{\fontfamily{pcr}\selectfont TCAF+BKNPL} approaches. On MJD 58792.5, the disk temperature ($T_{in}$) was $\sim 0.61$ keV with photon indices ($\Gamma_1, \Gamma_2$) 
of $\sim$ 3.19 and 1.96 and a break energy ($E_{break}$) of $\sim 11.39$ keV. Like the only NICER data fitting for that particular date, these values are consistent. 
The accretion rates were high with ${\dot m}_d = 10.15 ~{\dot M}_{Edd}$ and ${\dot m}_h = 2.92 ~{\dot M}_{Edd}$ with a very weak shock with $R = 1.64$ at a location 
of $X_s = 31.76 ~r_s$. The broken power-law indices ($\Gamma_1, \Gamma_2$) for {\fontfamily{pcr}\selectfont TCAF+BKNPL} model were 3.68, 2.02 with break energy 
10.66 keV. For the observation on MJD 58800.6, the $T_{in}$ decreased a little to $0.57$ keV with $\Gamma_1, \Gamma_2$ changed to 3.09, 2.07 and $E_{break} = 
9.44$ keV. The total accretion rate decreased with ${\dot m}_d$ decreasing to 6.06 and ${\dot m}_h$ increasing to 1.71 ${\dot M}_{Edd}$. The shock was almost the 
same compared to the last reported day with $R = 1.78$ and $X_s = 32.39$. $\Gamma_1, \Gamma_2$ was 3.45, 2.17 with $E_{break} = 9.01$ keV. On the next epoch on MJD 
58812, $T_{in}$ decreased a little more to $0.55$ keV with $\Gamma_1, \Gamma_2$ being 3.25, 1.92 and $E_{break} = 10.81$ keV. The total rate decreased further with 
${\dot m}_d$ and ${\dot m}_h$ decreasing to 4.86 and 1.61 ${\dot M}_{Edd}$. $X_s$ was $32.21 ~r_s$ with $R = 1.64$. $\Gamma_1, \Gamma_2$ were 3.64, 3.42 with 
$E_{break} = 10.18$ keV. 

This model describes the Comptonization of soft photons by the bulk motion of relativistic matter very close to the black hole. Illumination of the infalling 
matter at a close distance from the event horizon by the thermal radiations coming from the inner regions of the accretion disk is considered (Borozdin et al. 
For the broadband fitting, we notice that there is an extra feature for which we need an extra {\fontfamily{pcr}\selectfont power-law} component in terms of 
{\fontfamily{pcr}\selectfont broken power-law}. The slope of the spectra in the broadband range is changing at a break energy of $E_{break} \sim 9-11$ keV, which 
was not revealed in the $0.7-10$ keV energy band. Even for only the {\fontfamily{pcr}\selectfont DBB+PL} model fitting in the $0.7-10$ keV, there is a slight 
indication of a change in spectral slope. However, since it is not contributing to the residual due to the constraint of this energy range, only the 
{\fontfamily{pcr}\selectfont power-law} model was sufficient to fit the spectra for a good statistical value. This change of spectral shape at this energy is 
interesting as the source is in the HSS. We know that in the case of HSS, there could be a presence of bulk motion Comptonization that is observed to extend the 
spectrum up to very high energies. Also, this component is observed to contribute to the high energy band where, in HSS, thermal Comptonization is not dominant. 
Thus, we aimed to apply the \href{https://heasarc.gsfc.nasa.gov/xanadu/xspec/manual/node143.html}{\it bmc} model fitting also for our analysis. This model has 
four parameters: a characteristic thermal soft photon source black-body temperature ($kT$ in  keV, a spectral index ($\alpha$, where, 
$\Gamma_{bmc} = \alpha +1$), an illumination parameter ($A$; here it is taken as $log A$) that describes the fractional illumination of the infalling flow by 
the soft photon source, and the normalization ($N_{bmc}$). Here, $f=A/(A+1)$ (Borozdin et al. 1999), which is the fraction of the soft disk photons that 
are Comptonized in the process of bulk motion. Also, $N_{bmc} = L_{39}/d_{10}^2$, where $L_{39}$ and $d_{10}$ are the source luminosity in units of $10^{39}~ 
erg ~s^{-1}$ and source distance in units of $10$ kpc respectively. This is a self-consistent convolution model that is not an additive combination of power-law 
and thermal photon sources. Fitting by this model returns us good values of $\chi^2/DOF$ for both phenomenological and physical analysis cases. At first, when we 
fitted with the {\fontfamily{pcr}\selectfont constant*tbabs*(diskbb+power-law+bmc)} model, we link the parameter $kT$ of the BMC model with the $T_{in}$ of the 
disk black body model and found the best-fitted values of the other parameters. Then, fitting with the {\fontfamily{pcr}\selectfont constant*tbabs*(TCAF+power-law+bmc)}
model, we froze all four parameters of {\fontfamily{pcr}\selectfont bmc} model to the previous model fitted values and found the best fit. Thus, for the 
three obs. Ids. in HSS (on MJD 58792.5, 58800.6, and 58812), the seed photon temperature from {\fontfamily{pcr}\selectfont bmc} model fitting was $0.62$, $0.58$, 
and $0.56$ keV respectively. We found spectral indices of 0.14, 0.36, and 0.31, respectively, which make $\Gamma_{bmc}$ as 1.14, 1.36, and 1.31 for those three 
observations, respectively. The $log(A)$ has values of 0.41, 0.96, and 0.17, respectively, which make the Comptonization fraction due to BMC $\sim$ 0.7, 0.9, 
and 0.6, respectively. The normalizations are listed in Table 2. In Figure 5(a--b) we show the {\fontfamily{pcr}\selectfont constant*tbabs*(diskbb+power-law+bmc)} 
and {\fontfamily{pcr}\selectfont constant*tbabs*(TCAF+power-law+bmc)} model fitted spectra for the observation on MJD 58792.5. The fittings returned $\chi^2/DOF 
=$ 1.168 and 1.202 for the two models, respectively. On MJD 58800.6, the fittings returned $\chi^2/DOF$ values as 1.09, 1.06, and on MJD 58812 returned values 
1.29, 1.16, respectively. Changing from the model TCAF+BKNPL to TCAF+PL+BMC, the parameters of the TCAF model have changed only by a very small margin. 
The parameters are listed in Table 2. 

For the other 3 obs. ids outside the HSS phase, we did not need a broken power-law component. Only {\fontfamily{pcr}\selectfont DBB+PL} and 
{\fontfamily{pcr}\selectfont TCAF} models fitted these 3 observations to a very good statistical limit. On MJD 58866, $T_{in}$ decreased to a value to $0.17$ keV 
with $\Gamma$ being 1.81. On this day, we observed a very low disk rate of ${\dot m}_d = 0.04 ~{\dot M}_{Edd}$ with a low halo rate also as ${\dot m}_h 
= 0.04 ~{\dot M}_{Edd}$. The shock moved outwards and was at a location of $X_s = 111.05 ~r_s$ with $R = 1.41$. This supported the existence of the HIMS state 
during this epoch. On MJD 58879.6, $T_{in}$ was $0.22$ keV with $\Gamma = 1.67$ and the shock moved farther as $X_s$ was at $168.6 ~r_s$ with $R = 2.19$. Both 
the rates decreased further as disk rate became $0.007 ~{\dot M}_{Edd}$ and halo rate became $0.26 ~{\dot M}_{Edd}$. This epoch results support that the source 
was in HS in a declining phase. Although this combined fitting was achieved with only the power-law model (without disk black body), adding the disk blackbody 
component improved the fitting significantly from $\chi^2/DOF = 1.42$ to $\chi^2/DOF = 1.12$. Thus we reported the {\fontfamily{pcr}\selectfont DBB+PL} fit result 
for this epoch. For MJD 58889.6, $T_{in}$ was $0.27$ keV with $\Gamma = 1.64$. The shock moved much more outwards at $X_s = 246.9 ~r_s$ and shock strength was 
$R = 2.45$. ${\dot m}_d$ was $0.004 ~{\dot M}_{Edd}$ and ${\dot m}_h$ was $0.27 ~{\dot M}_{Edd}$ for this epoch. These results support that the source continued 
in declining HS before going into quiescence. For this epoch, we observed that only the power-law model could fit the data in the phenomenological analysis with 
$\chi^2/DOF = 1.22$. For only the {\fontfamily{pcr}\selectfont PL} model fitting, we found that $\Gamma$ had a value of 1.73 with no disk component. For this, we 
also checked the only NICER data fitting in the $0.7-10$ keV energy band. From there, also we found that for only the {\fontfamily{pcr}\selectfont PL} model 
fitting, $\Gamma = 1.69$, which is consistent with the broadband result. We also checked the TCAF fitting in the $0.7-10$ keV energy range also for this NICER 
observation data. We found similar result with ${\dot m}_d = 0.004 ~{\dot M}_{Edd}$ and ${\dot m}_h = 0.27 ~{\dot M}_{Edd}$ with a shock at $X_s = 246.8 ~r_s$ 
and $R = 2.41$. This shows the consistency of our analysis in both energy bands. 

While fitting with the phenomenological and physical approaches, we kept the hydrogen column density ($n_H$) parameter as free. In HSS, for DBB+BKNPL model fitting, 
$n_H$ was $0.069 \pm 0.017$, $0.092 \pm 0.0079$, and $0.07 \pm 0.0082$ $cm^{-2}$ for E1, E2, and E3 epochs respectively. Then, to fit with the 
{\fontfamily{pcr}\selectfont DBB+PL+BMC} model, we froze $n_H$ to those corresponding values for the 3 epochs respectively. Then again for {\fontfamily{pcr}\selectfont 
TCAF+BKNPL} model fit, we made the $n_H$ free and got $0.21 \pm 0.019$, $0.18 \pm 0.016$, and $0.12 \pm 0.039$ $cm^{-2}$ for E1, E2, and E3 respectively. Then we 
kept the $n_H$ frozen at these values for the {\fontfamily{pcr}\selectfont TCAF+PL+BMC} model fitting for 3 epochs. For epoch E4, the data is from XRT-NuSTAR. For 
this only, the Power-law model was not enough to achieve the best fit. For {\fontfamily{pcr}\selectfont DBB+PL}, and {\fontfamily{pcr}\selectfont TCAF} model 
fittings, we got $n_H =0.05 \pm 0.006$ and $0.06 \pm 0.001 ~cm^{-2}$ respectively. The E5, and E6 epochs, were also fitted using the only NICER model. From those 
fittings, we observed that all of the observations returned almost constant $n_H$ values for both the {\fontfamily{pcr}\selectfont DBB+PL} and 
{\fontfamily{pcr}\selectfont TCAF} models. From those analyses, we found the most probable $n_H = 0.056_{-0.006}^{+0.014} \times 10^{22} ~cm^{-2}$. Thus, while 
fitting for the broadband 
data, we froze $n_H$ to this most probable value for both these epochs and got best-fit results. From {\fontfamily{pcr}\selectfont TCAF} model fitting, we also got an 
estimation of the mass of this source in the range of $5.4-9.4 ~M_\odot$ with the average mass of $8.1_{-2.7}^{+1.3} ~M_\odot$.

\begin{table*}
\small
 \addtolength{\tabcolsep}{-1.5pt}
 \centering
	\caption{Spectral fitting results for high soft state (HSS) epochs}
 \label{tab:table2}
 \begin{tabular}{|c|c|c|c|c|c|c|c|}
 \hline
 Models  &   Parameters    &         E1$^{[1]}$        &          E2$^{[1]}$        &          E3$^{[1]}$         \\
\hline                                                           
         &     $n_H$       &   $6.9E-2  \pm 1.7E-2 $  &    $9.2E-2  \pm 7.9E-3 $   &    $7.0E-2  \pm 8.2E-3 $     \\
         &    $T_{in}$     &   $0.61    \pm 8.5E-4 $  &    $0.57    \pm 1.1E-3 $   &    $0.55    \pm 3.0E-3 $     \\
 DBB	    &     Norm        &   $2982.6  \pm 19.4   $  &    $2676.7  \pm 24.7   $   &    $1114.1  \pm 32.1   $     \\
 $+$     &    $\Gamma_1$   &   $3.19    \pm 1.3E-2 $  &    $3.09    \pm 9.0E-3 $   &    $3.25    \pm 3.4E-2 $     \\
 BKNPL   &   $E_{break}$   &   $11.39   \pm 0.16   $  &    $9.44    \pm 8.1E-2 $   &    $10.81   \pm 0.19   $     \\
         &    $\Gamma_2$	  &   $1.96    \pm 2.7E-2 $  &    $2.07    \pm 1.2E-2 $   &    $1.92    \pm 3.1E-2 $     \\
         &     Norm        &   $0.68    \pm 6.6E-3 $  &    $0.89    \pm 7.2E-3 $   &    $0.19    \pm 8.7E-3 $     \\
         & $\chi^2/DOF$    &   $     1066.7/909    $  &    $    1154.6/1204    $   &    $      877.6/695    $     \\
\hline                                                                       
         &     $n_H$       &   $      6.9E-2       $  &    $      9.2E-2       $   &    $       7.0E-2      $     \\
         &    $T_{in}$     &   $0.62    \pm 8.1E-4 $  &    $0.58    \pm 6.8E-4 $   &    $0.56    \pm 2.8E-3 $     \\
 DBB	    &     Norm        &   $2979.6  \pm 18.2   $  &    $2751.7  \pm 14.4   $   &    $1034.2  \pm 25.8   $     \\
 $+$     &    $\Gamma$ 	  &   $3.32    \pm 1.3E-2 $  &    $3.02    \pm 7.5E-3 $   &    $3.56    \pm 2.8E-2 $     \\
 PL	    &     Norm        &   $0.64    \pm 6.0E-3 $  &    $0.63    \pm 3.3E-3 $   &    $0.19    \pm 6.4E-3 $     \\
 $+$     & $KT$ (in $keV$) &   $     = T_{in}      $  &    $     = T_{in}      $   &    $     = T_{in}      $     \\
 BMC     &    $\alpha$     &   $0.14    \pm 4.5E-2 $  &    $0.36    \pm 2.5E-2 $   &    $0.31    \pm 4.9E-2 $     \\ 
         &     $log A$     &   $0.41    \pm 4.0E-2 $  &    $0.96    \pm 8.4E-2 $   &    $0.17    \pm 3.5E-2 $     \\
         &      Norm       &   $4.5E-4  \pm 9.1E-5 $  &    $6.3E-4  \pm 3.4E-5 $   &    $1.2E-4  \pm 7.3E-5 $     \\
         &  $\chi^2/DOF$   &   $    1060.7/908     $  &    $     1317.9/1203   $   &    $     897.7/694     $     \\
\hline                                                                        
         &     $n_H$       &   $0.21    \pm  1.9E-2$  &    $0.18    \pm  1.6E-2$   &    $0.12  \pm  3.9E-2  $     \\
         & ${\dot m}_d$    &   $10.15   \pm  0.05  $  &    $6.06    \pm  0.06  $   &    $4.86  \pm  0.24    $     \\
         & ${\dot m}_h$    &   $2.92    \pm  0.45  $  &    $1.71    \pm  0.14  $   &    $1.61  \pm  0.10    $     \\
 TCAF    &    $X_s$        &   $31.76   \pm  0.02  $  &    $32.39   \pm  0.01  $   &    $32.21 \pm  0.10    $     \\
 $+$	    &     $R$         &   $1.64    \pm  0.44  $  &    $1.78    \pm  0.02  $   &    $1.64  \pm  0.42    $     \\
 BKNPL   &  $\Gamma_1$	  &   $3.68    \pm  4.2E-2$  &    $3.45    \pm  3.5E-2$   &    $3.64  \pm  0.43    $     \\
         &  $E_{break}$    &   $10.66   \pm  0.17  $  &    $9.01    \pm  8.5E-2$   &    $10.18 \pm  2.58    $     \\
         &  $\Gamma_2$	  &   $2.02    \pm  0.10  $  &    $2.17    \pm  5.2E-2$   &    $3.43  \pm  0.56    $     \\
         &     Norm        &   $2.46    \pm  0.11  $  &    $1.89    \pm  0.10  $   &    $0.25  \pm  0.23    $     \\
         & $\chi^2/DOF$    &   $    1080.9/905     $  &    $     1214.04/1200  $   &    $   795.2/690       $     \\
\hline                                                                        
         &     $n_H$       &   $       0.21        $  &    $        0.18       $   &    $      0.12         $     \\
         & ${\dot m}_d$    &   $10.63   \pm 7.7E-2 $  &    $6.26    \pm  3.5E-2$   &    $4.84  \pm 0.13     $     \\
 TCAF    & ${\dot m}_h$    &   $3.34    \pm 2.6E-2 $  &    $2.34    \pm  5.8E-2$   &    $2.42  \pm 9.2E-2   $     \\ 
 $+$     &    $X_s$        &   $32.97   \pm 4.3E-2 $  &    $33.02   \pm  2.6E-2$   &    $30.9  \pm 3.93     $     \\ 
 PL      &     $R$         &   $1.89    \pm 0.63   $  &    $1.42    \pm  7.2E-2$   &    $1.57  \pm 0.49     $     \\ 
 $+$     &    $\Gamma$ 	  &   $3.49    \pm 5.7E-2 $  &    $3.21    \pm  3.0E-2$   &    $3.82  \pm 5.4E-2   $     \\
 BMC	    &     Norm        &   $1.30    \pm 0.12   $  &    $1.06    \pm  5.9E-2$   &    $0.46  \pm 4.1E-2   $     \\
         & $KT$ (in $keV$) &   $         0.62      $  &    $         0.58      $   &    $      0.56         $     \\
         &    $\alpha$     &   $         0.14      $  &    $         0.36      $   &    $      0.31         $     \\
         &    $log A$      &   $         0.41      $  &    $         0.96      $   &    $      0.17         $     \\
         &     Norm        &   $5.8E-4  \pm 8.6E-5 $  &    $6.3E-4  \pm  3.5E-5$   &    $1.2E-4\pm 4.3E-5   $     \\
         &  $\chi^2/DOF$   &   $    1188.9/906     $  &    $    1315.5/1202    $   &    $     803/693       $     \\
\hline
 \end{tabular}
 \noindent{
	 \leftline{$^{[1]}$ E1, E2, and E3 represent the epochs on MJD 58792.5, 58800.6, and 58812, respectively.}
	 \leftline{From {\fontfamily{pcr}\selectfont DBB+BKNPL} model fitting we have frozen the $n_H$ parameter for {\fontfamily{pcr}\selectfont DBB+PL+BMC} 
	 fitting and from {\fontfamily{pcr}\selectfont TCAF+BKNPL} model fitting we have frozen} 
	 \leftline{the $n_H$ parameter for {\fontfamily{pcr}\selectfont TCAF+PL+BMC} fitting for all 3 epochs respectively.}
	 \leftline{$n_H$ is in the units of $10^{22} cm^{-2}$.}
	 \leftline{We have also used the {\fontfamily{pcr}\selectfont tbnew\_feo} model as an alternate absorption model. The fit statistics are shown in the 
		   Appendix.}
          }
\end{table*}

\begin{table*}
 \addtolength{\tabcolsep}{-1.5pt}
 \centering
\caption{Spectral fitting results for other than HSS epochs}
 \label{tab:table3}
 \begin{tabular}{|c|c|c|c|c|c|c|c|}
 \hline
 Models  &   Parameters    &         E4$^{[1]}$        &          E5$^{[1]}$        &          E6$^{[1]}$         \\
\hline                                                 
         &     $n_H$       &              --           &    $      5.6E-2       $   &    $     5.6E-2        $    \\
 PL      &    $\Gamma$ 	  &              --           &    $1.79    \pm  8.1E-3$   &    $1.73    \pm  1.6E-2$    \\
         &     Norm        &              --           &    $8.0E-3  \pm  4.8E-5$   &    $4.1E-3  \pm  5.0E-5$    \\
         & $\chi^2/DOF$    &              --           &    $    1270.9/897     $   &    $    549.7/449      $    \\
\hline                                                                             
         &     $n_H$       &    $5.0E-2\pm 6.2E-3$     &    $     5.6E-2        $   &    $       5.6E-2      $    \\
 DBB	    &    $T_{in}$     &    $0.17  \pm 6.9E-3$     &    $0.22  \pm 1.3E-2   $   &    $0.27  \pm 4.2E-2   $    \\
 $+$	    &     Norm        &    $2.6E+4\pm 6932.5$     &    $183.5 \pm 54.5     $   &    $26.25 \pm 17.85    $    \\
 PL      &    $\Gamma$     &    $1.81  \pm 1.0E-2$     &    $1.67  \pm 1.1E-2   $   &    $1.64  \pm 2.3E-2   $    \\
         &     Norm        &    $3.0E-2\pm 1.1E-3$     &    $6.7E-3\pm 1.1E-4   $   &    $3.5E-3\pm 1.3E-4   $    \\	 
         & $\chi^2/DOF$    &    $   689.2/689    $     &    $    1003.1/895     $   &    $     519/447       $    \\
\hline                                                                             
\hline                                                                             
         &     $n_H$       &    $5.6E-2 \pm 1.0E-3 $   &    $     5.6E-2        $   &    $      5.6E-2       $    \\
         & ${\dot m}_d$    &    $3.8E-2 \pm 2.2E-2 $   &    $6.7E-3  \pm  4.0E-3$   &    $5.6E-2  \pm  3.7E-3$    \\
TCAF	 & ${\dot m}_h$    &    $0.47   \pm 0.003  $   &    $0.26    \pm  0.004 $   &    $0.27    \pm  0.01  $    \\
         &    $X_s$        &    $111.1  \pm 1.4    $   &    $168.6   \pm  1.01  $   &    $246.9   \pm  3.8   $    \\
         &     $R$         &    $1.41   \pm 1.0E-3 $   &    $2.19    \pm  0.005 $   &    $2.45    \pm  0.008 $    \\
         & $\chi^2/DOF$    &    $  700.7/687       $   &    $    1257.7/893     $   &    $     563/445       $    \\
\hline
 \end{tabular}
 \noindent{
	 \leftline{$^{[1]}$ E4, E5, and E6 represent the epochs on MJD 58866, 58879.6, and 58889.6, respectively.}
	 \leftline{For E5 and E6, we have frozen the $n_H$ value to $0.056 ~cm^{-2}$ as from the NICER data fitting, we found this as the average of $n_H$ when 
	 the source} \leftline{was in declining HS. $n_H$ is in the units of $10^{22} cm^{-2}$.}
          }
\end{table*}

\section{Discussions} 

The Galactic black hole candidate MAXI J0637-430 was monitored on a daily basis by the Neutron star Interior Composition ExploreR (NICER) satellite from November 
2019 to May 2020. The Swift/XRT also monitored this period on a daily basis. Looking at the outburst light curve, we notice that the outburst is a fast rise slow 
decay (FRSD) type which lasted for a very short time in the rising phase. The source made a very fast transition to its thermal dominant soft state after the onset 
of the outburst. We have used archival NICER, XRT, and NuSTAR data to study the broadband spectroscopy of this source on 6 different epochs during the 2019--20 outburst. 
Using combined NICER+NuSTAR (4 obs.) and XRT+NuSTAR (2 obs.) data, we have spectrally fitted the outburst in the $0.7-70$ keV energy band. However, we first checked 
spectral fitting using only the NICER data in the $0.7-10$ keV energy band for the entire duration of the outburst. Our spectral fitting was done using two 
approaches, i.e., with phenomenological and physical models. The $0.7-10$ keV NICER data were fitted using the {\fontfamily{pcr}\selectfont DBB+PL}, 
{\fontfamily{pcr}\selectfont TCAF+PL} (for some observations) and only {\fontfamily{pcr}\selectfont TCAF} (for other observations) models. However, for 
broadband fitting (in $0.7-70$ keV) in 3 epochs, the {\fontfamily{pcr}\selectfont DBB+PL} model was unable to fit the spectra for a good $\chi^2/DOF$ value. The 
same was also with the {\fontfamily{pcr}\selectfont TCAF} model fitting. We could not fit the first 3 epochs properly with only the {\fontfamily{pcr}\selectfont 
TCAF+PL} model. We use {\fontfamily{pcr}\selectfont DBB+BKNPL} and {\fontfamily{pcr}\selectfont TCAF+BKNPL} models to fit these 3 epochs of data to obtain a good 
$\chi^2/DOF$ value. The other 3 epochs were fitted with {\fontfamily{pcr}\selectfont DBB+PL} and {\fontfamily{pcr}\selectfont TCAF} models. Since reporting this 
evolution is not the aim of this work, we report only those 6 epochs for which we perform the broadband analysis. 

According to the previous reports (Jana et al. 2021 and the references therein), the first three out of our analyzed epochs belong to HSS, while one is in 
HIMS and the other two are in declining HS. We notice that, in the HSS phase, the data is better fitted with the phenomenological approach compared to the 
physical approach. When we fitted with the TCAF model, there is residual present in the low energy (Fig 2c, 4b, 5b), which are not prominently noticeable when fitting 
with the various combination of phenomenological models (Fig. 2a, 3b, 5a). It could be due to contributions from two effects. The NICER data suffers from residuals in lower
energies due to absorption edge features of Si in $1.7-2.1$ keV, Au in $2.2-2.3$ keV, etc., due to instrumental error and/or error in the response function. Along
with this, we generally found that it was harder to fit spectra with the TCAF model when a source was in the soft states (SS or HSS). Combinedly, these can contribute to 
the presence of residual in the lower energies in HSS. It is the same for the XRT observations also. In case of XRT, the spectrum suffers from high residuals at energies
$<5$ keV due to an error in the response function. From our spectral analysis, we find that the inner-disk temperature ($T_{in}$) was the highest for the first epoch and 
decreased by a small value over the next two epochs. From the {\fontfamily{pcr}\selectfont DBB+BKNPL} fitting, we also noticed that the $\Gamma_1$ was $>3$ during 
these epochs. From physical model analysis results, we found very high accretion rates (disk rate dominant over halo rate) and the presence of weak shock. During these 
3 epochs, the shock was very close to the compact black hole ($X_s \sim 31-33 ~r_s$), which actually reflects the fact that thermal radiation became dominant. For all 
these 3 epochs, we see that there was the dominance of ${\dot m}_d$ over ${\dot m}_h$, although both of these rates decreased gradually, suggesting that the source 
had already reached its peak and was in the declining phase. From these results, these observations agree with the previous findings that the source was in HSS. For 
the next 3 epochs, $T_{in}$ decreased to very low values with lower photon indices. The accretion rates also decreased to very low values, and the shock gradually 
went outwards and became strong. This suggests that the disk photons were unable to cool the CENBOL region and were at a far distance from the source, which is due 
to less supply of Keplerian photons. This also agrees with the previous findings that the source was in HIMS and HS during the fourth and fifth-sixth observations, 
respectively.

While fitting with only NICER data, there was a requirement of a higher value of photon index of power-law ($\Gamma$) to achieve good fitting statistics (Fig. 
2a). This was the case for all the NICER observations present in the HSS. In this phase, the disk was very close to the compact object. The 
{\fontfamily{pcr}\selectfont TCAF} model was not able to fit the spectral data with acceptable statistics (Fig. 2b). This difficulty in fitting the data with the 
{\fontfamily{pcr}\selectfont TCAF} model in the soft state was also encountered by our group for sources like XTE J1752-223, MAXI J1910-057, etc. However, we could 
achieve an acceptable best fit for these sources using only the {\fontfamily{pcr}\selectfont TCAF} model. For this source, however, it was not the case. For the 
$0.7-10$ keV energy band, we needed an extra component {\fontfamily{pcr}\selectfont power-law} here) to achieve an acceptable best fit. To understand the requirement 
of this extra component in a better way, we decided to check the spectral fitting in a broader energy band. Hence we took combined NICER+NuSTAR and XRT+NuSTAR data 
in the $0.7-70$ keV energy band. We found that the {\fontfamily{pcr}\selectfont DBB+PL} and also the {\fontfamily{pcr}\selectfont TCAF+PL} models were unable to 
fit the spectra (Fig. 3a \& 4a). However, implementing a {\fontfamily{pcr}\selectfont broken power-law} in place of a {\fontfamily{pcr}\selectfont power-law} 
improves the fitting significantly. We noticed from the unfolded spectrum fit that there is a different slope of the spectrum at around $9-11$ keV. This is also 
listed in Table 2. This spectral break is seen only for the observations present in HSS. We suspect that since, in this state, the inner-disk is very close to the 
compact object, and the CENBOL size is also very small, there could be the presence of bulk motion Comptonization. There is an advection of matter from the inner 
regions to the compact object. At regions very near the BH, advective matter falls freely with relativistic speed. Since the bulk of matter is moving with this 
speed, photons are Comptonized and upscattered due to this bulk motion. The matter is cold in these inner regions, and this Comptonization is not thermal in nature. 
Thus this radiation is very negligible in lower energies, where the thermal blackbody radiation and thermal Comptonization are very dominant. This upscattering of 
photons by bulk motion contributes to the higher energy end when the contributions from blackbody radiation decrease almost entirely. Since the source is in HSS, 
the thermal Comptonization is also not prominent in nature. For this, the contribution from the BMC can be prominently observed at higher energies. This effect has 
changed the spectral slope and extended the spectrum to very high energies. The spectral break at around $9-11$ keV energy range in our cases signifies where the 
BMC has started to contribute significantly. Titarchuk et al. (2009) reported the presence of bulk motion Comptonization phenomenon in the black hole candidate GRS 
1915+105. For HSS, they made use of two {\fontfamily{pcr}\selectfont bmc} components to account for the soft thermal component and the hard component with a 
turnover at high energies. According to them, the hard component is due to the Comptonization of soft photons in the converging flow, very near to the BH.

To make our assumption stronger, i.e., the possible presence of the BMC phenomenon in this source, we fitted the 3 observations in HSS using the 
{\fontfamily{pcr}\selectfont bmc} model also. However, we first used the combined {\fontfamily{pcr}\selectfont diskbb + power-law + bmc} model to account for the 
contribution from the blackbody radiation, thermal Comptonization, and bulk motion Comptonization. The fitting is given in Fig. 5a. Then, we checked this model with 
the TCAF model. We observed that both these combined models fit the data well. This also solidifies our assumption that there could be the presence of bulk motion 
Comptonization in the HSS. We also checked the luminosity of the source during these different epochs in HSS. Comparing with the Eddington luminosity, we find that 
this source radiated in the sub-Eddington luminosity during these epochs, in which the luminosity was $0.043 ~L_{Edd}$, $0.03 ~L_{Edd}$, and $0.011 ~L_{Edd}$ 
respectively. From the BMC normalization ($N_{bmc}$), we also estimated that the BMC component contributes $0.0003 ~L_{Edd}$, $0.0004 ~L_{Edd}$, and $0.0001
~L_{Edd}$ to the source luminosity respectively. The contribution from the BMC is significantly less. This is because the BMC is dominant in the high-energy part 
of the spectrum, where blackbody radiation and thermal Comptonization do not contribute that much. However, in low energy, thermal radiation is much more 
dominant than the BMC. The $N_{bmc}$ is thus very low from the spectral fitting, which reflects this fact. BMC extends the tail of the spectrum to high energy 
without contributing that much in the low energy part, where there is the dominance of thermal radiation.

The hydrogen column density varied between $0.05-0.21 \times 10^{22} ~cm^{-2}$ during these 6 epochs with the average column density of $0.056^{+0.014}_{-0.006} 
\times 10^{22} ~cm^{-2}$. This range is due to the choice of a combination of models. Excluding the physical model fitting approach in the HSS, we found 
that $n_H$ varied in the range of $0.05-0.09 \times 10^{22} ~cm^{-2}$. However, the inclusion of the TCAF model in the HSS broadens this range, which explains 
our prior mention that it is harder to fit data in the HSS with the TCAF model. However, this is not outside the range of those previous reports (Ma et al. 2022; 
Lazar et al. 2021). For our spectral analysis, we found the mass to vary in the range of $5.4-9.4 ~M_\odot$ with the average mass of $8.1~M_\odot$. This
is in agreement with the previous reportings of Jana et al. (2021), Soria et al. (2022). 
This is not the actual variation of the mass of the source during the outburst. The mass of a black hole is non-changing during an outburst. However, mass varies 
as the fourth power of the disk temperature ($\sim T_{disk}^4$). Thus any error in measuring the $T_{disk}$ leads to a high error in the measurement of the mass 
of the source. Our finding of mass is in agreement with the previous reports.

\section{Summary and Conclusions}

We have studied the spectral properties of the BHC MAXI J0637-430 during its 2019--20 outburst. First, we showed the evolution of the light curve profile during the 
entire duration of the outburst and showed the variation of the hardness ratio during the entire outbursting phase. However, studying the evolution of the outburst 
was not the aim of our work. We wanted to study the broadband properties of this source. Simultaneous NuSTAR data was available on 4 days with NICER observations and 
on 2 days with XRT observations. Using those publicly available NICER+NuSTAR and XRT+NuSTAR data in the $0.7-70$ keV energy band, we spectrally fitted data in these 
6 epochs. We first tried to fit the data using only the {\fontfamily{pcr}\selectfont diskbb+power-law} and {\fontfamily{pcr}\selectfont TCAF} models. However, these 
models were unable to fit the data for the observations in the HSS phase. Other 3 obs. data were fitted properly with the above-mentioned models for acceptable 
statistics. For the first 3 epochs, we used {\fontfamily{pcr}\selectfont diskbb+broken power-law} and {\fontfamily{pcr}\selectfont TCAF+broken power-law} models. This 
set of models could fit the data well for an acceptable $\chi^2/DOF$ value. We also checked the fitting using {\fontfamily{pcr}\selectfont diskbb+power-law+bmc} and 
{\fontfamily{pcr}\selectfont TCAF+power-law+bmc} models for these 3 epochs. Based on our spectral fitting, we have come to the following conclusions.

i) The source was in a high soft state during the first 3 epochs of our studied period. In these epochs, the flux was very high, with the dominance of the thermal 
component over the non-thermal component. This was also supported by the dominance of disk rate over halo rate. The photon index of power-law was also very high, which 
solidifies the presence of HSS. During these epochs, we think that the disk was at a very close distance from the black hole. Due to the advection of matter from the 
inner regions to the source, matter falls freely with relativistic speed due to very high gravitation. Due to the motion of this bulk matter, photons are upscattered 
to high energies. This supports our assumption that BMC could be present during these 3 epochs of our studied period.

ii) For the other 3 epochs, the source was in the declining hard intermediate and hard states. Here both the rates decreased significantly, and the inner regions of 
the disk moved far away from the source. Viscosity was already turned off, which led to the outward movement of the CENBOL due to the non-effectiveness of the Compton 
cooling process.

iii) The hydrogen column density varied between $0.05-0.21 \times 10^{22} ~cm^{-2}$, and the mass of the source varied between $5.4-9.4 ~M_\odot$.

\section{Data Availability}

This work has made use of public data from several satellite/instrument archives and has made use of software from the HEASARC, which is developed and monitored by 
the Astrophysics Science Division at NASA/GSFC and the High Energy Astrophysics Division of the Smithsonian Astrophysical Observatory. XRT data is provided by the 
UK Swift Science Data Centre at the University of Leicester. NICER data is provided by NASA/GSFC. FPM data from the NuSTAR mission is led by Caltech, which is funded 
by NASA and managed by NASA/JPL. The NuSTARDAS software package is jointly developed by ASDC, Italy and Caltech, USA.

\section{Acknowledgements}

K.C. acknowledges visiting research grant (NSTC 111-2112-M-007-019) from the Institute of Astronomy, National Tsing Hua University (NTHU). D.D. acknowledges support 
from the NSTC project (grant number: 111-2811-M-007-066) of NTHU. S.K.N. acknowledges support from the SVMCM fellowship, the government of West Bengal. H.-K.C. 
acknowledges support from the NSTC project (grant number: 111-2112-M-007-019) of NTHU.

\section*{\Large \bf Appendix}
\vspace{0.3cm}

The tested models of our analysis showed residual in the lower energy region ($< 5$ keV), as we can observe from Fig. 2, 3, 4, 5. Our analyzed results 
showed a variation of $n_H$ in the range of ($0.05-0.2) \times 10^{22} ~cm^{-2}$. As suggested by the referee, we checked if the data in the HSS can be fitted, keeping 
the $n_H$ almost constant or in a very narrow range. We checked the fittings keeping $n_H$ fixed at the Galactic value of $0.052 \times 10^{22} ~cm^{-2}$ (HI4PI 
collaboration). We did not see any noticeable change in the lower energy residuals or improvement in the $\chi^2/DOF$ values in this method.
The fitted parameters were also in the error range of the case when fitted, keeping the $n_H$ free. For the physical approach (using the TCAF model), this was 
also the case. However, in this case, freezing the $n_H$ at the Galactic value actually increased the $\chi^2/DOF$ values for the observations. This also did 
not improve the lower energy residuals.

As previously mentioned in the data analysis section (section 2.3), we have reanalyzed the data using the {\fontfamily{pcr}\selectfont tbnew\_feo} model as 
the absorption model. Using this absorption model, we fitted all the previously used combinations of models. Taking reference from Lazar et al. (2021), we used 
subsolar values of $<0.6$ Solar. Although there were some changes in the $\chi^2/DOF$ values, the lower energy residuals were not improved by any significance.

In Table A, we show the fitted $\chi^2/DOF$ values when we used {\fontfamily{pcr}\selectfont tbabs} (both cases when $n_H$ is free and fixed) and 
{\fontfamily{pcr}\selectfont tbnew\_feo} as our absorption models for separate combinations of phenomenological and physical models for the three epochs in HSS.

\begin{table}[h!]
  \vskip 0.1cm
  \centering{Table A: $\chi^2/DOF$ values using different absorption models}
  \centering
  \vskip 0.2cm
 \addtolength{\tabcolsep}{-1.5pt}
 \begin{tabular}{|c|c|c|c|c|}
 \hline
   (1)   &    (2)      &             (3)            &               (4)             &                        (5)                            \\
 Models  &   Epochs    &         $n_H=$ free        &          $n_H=$ frozen        &     Using {\fontfamily{pcr}\selectfont tbnew\_feo}    \\
\hline
 DBB     &    E1       &         $1066.7/909$       &          $1063.9/909$         &                     $1066.8/908$                      \\
 $+$     &    E2       &         $1154.6/1204$      &          $1129.1/1204$        &                    $1361.95/1203$                     \\
 BKNPL   &    E3       &         $877.6/695$        &          $853.7/695$          &                     $858.7/695$                       \\
\hline
 DBB$+$  &    E1       &         $1060.7/908$       &          $1066.1/908$         &                     $1067.6/907$                      \\
 PL$+$   &    E2       &         $1317.9/1203$      &          $1398.8/1203$        &                    $1330.20/1203$                     \\
 BMC     &    E3       &         $897.7/694$        &          $893.3/694$          &                     $894.1/694$                       \\
\hline
 TCAF    &    E1       &         $1080.9/905$       &          $1536.7/905$         &                     $1134.4/905$                      \\
 $+$     &    E2       &         $1214.04/1200$     &          $1550.2/1200$        &                    $1222.9/1199$                      \\
 BKNPL   &    E3       &         $795.2/690$        &          $806.7/691$          &                     $792.5/690$                       \\
\hline
 TCAF$+$ &    E1       &         $1188.9/906$       &          $1776.6/906$         &                     $1143.6/905$                      \\
 PL$+$   &    E2       &         $1315.5/1202$      &          $1926.4/1202$        &                    $1309.9/1201$                      \\
 BMC     &    E3       &         $803/693$          &          $806.5/693$          &                     $798.2/693$                       \\
\hline
 \end{tabular}
  \vskip 0.1cm
\noindent{
         \leftline{Cloumn 1 represents the combination of models used for spectral analysis.}
         \leftline{Here, E1, E2, and E3 are epochs at MJD 58792.5, 58800.6, and 58812, respectively.}
         \leftline{In columns 3, \& 4, we show the $\chi^2/DOF$ values for different combinations of models for the 3 different epochs when we used
         {\fontfamily{pcr}\selectfont tbabs} model. In column 3,} \leftline{we show the values when $n_H$ was kept free, and in column 4, we show the
         values when $n_H$ was fixed at the Galactic value.}
         \leftline{In column 5, we show the $\chi^2/DOF$ values for the same when we replaced the {\fontfamily{pcr}\selectfont tbabs} model with the
         {\fontfamily{pcr}\selectfont tbnew\_feo} model.}
         }
\end{table}

\end{document}